\documentclass[aps, prd, amsmath, amssymb, amsfonts, floats, %
floatfix, superscriptaddress, nofootinbib, twocolumn, showpacs]%
{revtex4}

\allowdisplaybreaks[2]

\usepackage{graphicx}
\usepackage[usenames, dvipsnames]{color}
\usepackage[colorlinks, pdfborder={0 0 0}, plainpages=false]{hyperref}
\usepackage{breakurl}
\usepackage{amsmath, amssymb, amsfonts}
\usepackage{xspace} 
\usepackage{dcolumn}
\usepackage{bm}
\usepackage{multirow} 

\def\laq{\raise 0.4ex\hbox{$<$}\kern -0.8em\lower 0.62ex\hbox{$\sim$}}
\def\gaq{\raise 0.4ex\hbox{$>$}\kern -0.7em\lower 0.62ex\hbox{$\sim$}}

\definecolor{CiteColor}{rgb}{0, 0.5, 0} %
\hypersetup{citecolor=CiteColor} %
\definecolor{RefColor}{rgb}{0.55, 0, 0} %
\hypersetup{linkcolor=RefColor} %

\usepackage{ulem}
\definecolor {darkgreen}{rgb}{0.2, 0.7, 0.2}

\newcommand{\ab}[1]{\textcolor{RedViolet}{#1}}

\newcommand{\Guelph}{\affiliation{Department of Physics, University of
    Guelph, Guelph, ON N1G 2W1, Canada}}
\newcommand{\Caltech}{\affiliation{Theoretical Astrophysics 350-17,
    California Institute of Technology, Pasadena, CA 91125, USA}}
\newcommand{\CITA}{\affiliation{Canadian Institute for Theoretical
    Astrophysics, 60 St. George Street,\\ University of Toronto,
    Toronto, ON M5S 3H8, Canada}}
\newcommand{\Cornell}{\affiliation{Center for Radiophysics and Space
    Research, Cornell University, Ithaca, NY, 14853, USA}}
\newcommand{\Maryland}{\affiliation{Maryland Center for Fundamental
    Physics \& Joint Space-Science Institute,\\ Department of Physics,
    University of Maryland, College Park, MD 20742, USA}}
\newcommand{\Radcliffe}{\affiliation{Radcliffe Institute for Advanced
    Study, Harvard University, 8 Garden St., Cambridge, MA 02138,
    USA}}

\newcommand{\MM}{\mathcal{M}}

\newcommand{\roughly}{\mathchar"5218\relax} 

\newcommand{\vP}{\mbox{\boldmath${P}$}}
\newcommand{\vS}{\mbox{\boldmath${S}$}}
\newcommand{\vR}{\mbox{\boldmath${R}$}}

\newcommand{\vvr}{\mbox{\boldmath${r}$}}
\newcommand{\vp}{\mbox{\boldmath${p}$}}

\newcommand{\vDelta}{\mbox{\boldmath${\Delta}$}}

\begin{document}

\title{Prototype effective-one-body model for nonprecessing
  spinning inspiral-merger-ringdown waveforms}

\author{Andrea Taracchini} \Maryland %
\author{Yi Pan} \Maryland %
\author{Alessandra Buonanno} \Maryland \Radcliffe%
\author{Enrico Barausse} \thanks{CITA National Fellow} \Guelph
\Maryland%
\author{Michael Boyle} \Cornell %
\author{Tony Chu} \CITA %
\author{Geoffrey Lovelace} \Cornell %
\author{Harald P. Pfeiffer} \CITA %
\author{Mark A. Scheel} \Caltech %

\begin{abstract}
  This paper presents a tunable effective-one-body (EOB) model for
  black-hole (BH) binaries of arbitrary mass ratio and aligned
  spins. This new EOB model incorporates recent results of
  small-mass-ratio simulations based on Teukolsky's perturbative
  formalism. The free parameters of the model are calibrated to
  numerical-relativity simulations of nonspinning BH-BH systems of
  five different mass ratios and to equal-mass non-precessing BH-BH systems with
  dimensionless BH spins $\chi_i\simeq\pm 0.44$.  The present analysis
  focuses on the orbital dynamics of the resulting EOB model, and on
  the dominant ($\ell$,$m$)=(2,2) gravitational-wave mode.
  The calibrated EOB model can generate inspiral-merger-ringdown
  waveforms for non-precessing, spinning BH binaries with any mass
  ratio and with individual BH spins $-1 \leq \chi_i \lesssim 0.7$.
  Extremizing only over time and phase shifts, the calibrated EOB
  model has overlaps larger than 0.997 with each of the seven
  numerical-relativity waveforms for total masses between $20M_\odot$
  and $200M_\odot$, using the Advanced LIGO noise curve.
  We compare the calibrated EOB model with two additional equal-mass
  highly spinning ($\chi_i \simeq -0.95, +0.97$) numerical-relativity
  waveforms, which were not used during calibration.  We find that the
  calibrated model has overlap larger than 0.995 with the simulation
  with nearly extremal \emph{anti-aligned} spins.  Extension of this
  model to black holes with \emph{aligned} spins $\chi_i \gtrsim 0.7$
  requires improvements of our modeling of the plunge dynamics and
  inclusion of higher-order PN spin terms in the gravitational-wave
  modes and radiation-reaction force.
\end{abstract}

\date{\today}


\maketitle

\section{Introduction}
\label{sec:intro}

Coalescing compact-object binary systems (binaries, for short) are
among the most promising sources of gravitational waves (GWs) for
detectors like the U.S. Laser Interferometer Gravitational-Wave
Observatory (LIGO), the British-German GEO, and the French-Italian
Virgo~\cite{Abbott:2007, Grote:2008zz, Acernese:2008}.  LIGO and Virgo
are undergoing upgrades to Advanced
configurations~\cite{Shoemaker2009}, which will improve sensitivity by
about a factor of 10. A detailed and accurate understanding of the GWs
radiated as the bodies in a binary spiral towards each other is
crucial not only for the initial detection of such sources, but also
for maximizing the information that can be obtained from the GW
signals once they are observed.

The matched-filtering technique is the primary data-analysis tool used
to extract the GW signals from the detectors' noise. It requires
accurate waveform models of the expected GW signals.  Analytical
templates based on the post-Newtonian (PN)
approximation~\cite{Sasaki:2003xr, Blanchet2006, Futamase:2007zz,
  Goldberger:2004jt} of the Einstein field equations developed over
the past thirty years accurately describe the inspiraling stage of the
binary evolution. In 1999 a new approach to the two-body dynamics of
compact objects, the so-called effective-one-body (EOB) approach, was
proposed with the goal of extending the analytical templates
throughout the last stages of inspiral, plunge, merger, and
ringdown. The EOB approach uses the results of PN theory, black-hole
perturbation theory, and, more recently, the gravitational self-force
formalism. It does not, however, use the PN results in their original
Taylor-expanded form (i.e., as polynomials in $v/c$), but in a
resummed form.

The EOB formalism was first proposed in Refs.~\cite{Buonanno99,
  Buonanno00} and subsequently improved in Refs.~\cite{DJS00,
  Damour01c, Buonanno06}. Using physical intuition and results from
black-hole perturbation theory and the close-limit approximation,
Refs.~\cite{Buonanno00, Buonanno06} computed preliminary plunge,
merger, and ringdown signals of nonspinning and spinning black-hole
binaries. After breakthroughs in numerical relativity
(NR)~\cite{Pretorius2005a, Baker2006a, Campanelli2006a}, the EOB
inspiral-merger-ringdown waveforms were improved by calibrating the
model to progressively more accurate NR simulations, spanning larger
regions of the parameter space~\cite{Buonanno-Cook-Pretorius:2007,
  Pan2007, Buonanno2007, Damour2007a, DN2007b, Boyle:2008, DN2008,
  Buonanno:2009qa, Pan:2009wj, Damour2009a, Pan:2011gk}. More
recently, an EOB model for the dominant $(2,2)$ mode and four
subdominant modes was built for nonspinning binaries of
comparable masses~\cite{Pan:2011gk} and the small-mass-ratio
limit~\cite{Barausse:2011kb}.  These results, at the interface between
numerical and analytical relativity, have already had an impact in
LIGO and Virgo searches. The first searches of high-mass and intermediate-mass 
black-hole binaries in LIGO/Virgo data~\cite{Abadie:2011kd,Abadie:2012} used the
inspiral-merger-ringdown templates generated by the EOB model
calibrated in Ref.~\cite{Buonanno2007}, as well as the
phenomenological templates proposed in Ref.~\cite{Ajith:2008}.

Stellar-mass black holes are expected to carry spins, which
significantly increases the dimension of the binary parameter
space. The first EOB Hamiltonian with leading-order (1.5PN) spin-orbit
and (2PN) spin-spin couplings was developed in
Ref.~\cite{Damour01c}. Then, Ref.~\cite{Buonanno06} worked out the
radiation-reaction force in the EOB equations of motion in the
presence of spins and computed inspiral-merger-ringdown waveforms for
generic spinning binaries, capturing their main features, including
the so-called ``hang up''. Later, Ref.~\cite{Damour:024009}
incorporated the next-to-leading-order (2.5PN) spin-orbit couplings in
the EOB Hamiltonian. By construction, in the test-particle limit the
Hamiltonian of Ref.~\cite{Damour:024009} does not reduce to the
Hamiltonian of a spinning test particle in the Kerr spacetime.
Moreover, the Hamiltonian of Ref.~\cite{Damour:024009} rewrites the
EOB radial potential using Pad\'e summation, causing spurious poles in
some regions of parameter space. Nevertheless, the Hamiltonian of
Ref.~\cite{Damour:024009} was adopted in Ref.~\cite{Pan:2009wj} to
demonstrate the possibility of calibrating the EOB model for spinning
binaries.

Since then, substantial progress has been made towards improving the
spin EOB Hamiltonian. Ref.~\cite{Barausse:2009aa} worked out the
Hamiltonian for a spinning test-particle in a generic spacetime, which
was used in Ref.~\cite{Barausse:2009xi} to derive a spin EOB
Hamiltonian having the correct test-particle limit. Furthermore,
Ref.~\cite{Barausse:2009xi} rewrote the EOB radial potential in a way
that guarantees the absence of poles without employing the Pad\'e
summation. As a consequence, the EOB Hamiltonian of
Ref.~\cite{Barausse:2009xi} has desirable strong-field circular-orbit
features, such as the existence of an innermost-stable circular orbit
(ISCO), a photon circular orbit (or light-ring), and a maximum in the
orbital frequency during the plunge. Still preserving these
properties, the spin EOB Hamiltonian of Ref.~\cite{Barausse:2009xi}
was recently extended to include the next-to-next-to-leading-order
(3.5PN) spin-orbit couplings in Ref.~\cite{Barausse:2011ys}. The EOB
Hamiltonian of Ref.~\cite{Damour:024009} was also recently extended
through 3.5PN order in the spin-orbit sector in
Ref.~\cite{Nagar:2011fx}.

In the non-conservative sector of the EOB model, the
radiation-reaction force in the EOB equations of motion is built from
the GW energy flux, which, in turn, is computed from a decomposition
of the waveform into spherical harmonic $(\ell, m)$ modes. These
modes, instead of being used in their Taylor-expanded form, are
resummed (or factorized).  This factorization was originally proposed
in Refs.~\cite{Damour2007, DIN} for nonspinning black-hole binaries,
and was then extended to include spin effects in Ref.~\cite{Pan2010hz}
and higher-order PN spinless terms in Refs.~\cite{Fujita:2010xj,
  Fujita:2011zk}. In the test-particle limit, the factorized waveforms
are known at very high PN order---for example their sum generates the
GW energy flux for nonspinning binaries through
14PN~\cite{Fujita:2011zk} order and to 4PN order in terms involving
the black-hole spins.  However, in the comparable-mass case the GW
modes are known only at a much lower PN order. Despite the fact that
the GW energy flux in the comparable-mass case is known through
3.5PN~\cite{Kidder2008, BFIS} and 3PN~\cite{Blanchet:2011zv} order in
the nonspinning and spin-orbit sectors, and 2PN order in the
spin-spin sector, the GW modes have been computed only through 1.5PN
order for spin-orbit couplings and 2PN order for spin-spin
couplings~\cite{Arun:2009, Pan2010hz}.  Currently, this lack of
information in the GW modes is the main limitation of our spin EOB
model, and, as we will see, it affects the performance of the model
for prograde orbits and large spin values.

In this paper, we build upon the past success in analytically modeling
inspiral-merger-ringdown waveforms through the EOB formalism, and
develop a prototype EOB model for non-precessing spinning black-hole
binaries that covers a large region of the parameter space and can be
used for detection purposes and future calibrations. More
specifically, we adopt the EOB Hamiltonian derived in
Refs.~\cite{Barausse:2009xi, Barausse:2011ys}, the GW energy flux and
factorized waveforms derived in Refs.~\cite{DIN, Pan2010hz}, and
calibrate the EOB (2,2) dominant mode to seven NR waveforms: five
nonspinning waveforms with mass ratios $1,1/2,1/3,1/4$ and
$1/6$~\cite{Pan:2011gk} and two equal-mass non-precessing spinning
waveforms of spin magnitudes $0.44$~\cite{Chu2009}.  We combine the
above results with recent small-mass-ratio results produced by the
Teukolsky equation~\cite{Barausse:2011kb} to build a prototype EOB
model for inspiral-merger-ringdown waveforms for non-precessing
spinning black-hole binaries with any mass ratio and individual
black-hole spins $-1 \leq \chi_i \lesssim 0.7$. For $\chi_i \gtrsim
0.7$, although the EOB dynamics can be evolved until the end of the
plunge, the EOB (2,2) mode peaks too early in the evolution, where the
motion is still quasicircular. As a consequence, we cannot correct the
EOB (2,2) mode to agree with the NR (2,2) mode peak using
non-quasicircular amplitude coefficients. This limitation, which also
affects the small-mass-ratio limit results~\cite{Barausse:2011kb}, is
caused by the poor knowledge of PN spin effects in the GW modes and
makes the prototype EOB waveforms unreliable for $\chi_i \gtrsim
0.7$. Two NR waveforms with nearly extremal spin
magnitudes~\cite{Lovelace2010, Lovelace:2011nu} became available to us
when we were finishing calibration of the spin EOB model. We use them
to examine the limitations of the spin prototype EOB model, and
extract from them useful information for future work.

The paper is organized as follows. In Sec.~\ref{sec:EOB-model}, we
describe the spin EOB model used in this work, its dynamics,
waveforms, and adjustable parameters.  Section~\ref{sec:EOB-cal}
discusses the alignment procedure used to compare EOB and NR waveforms
at low frequency, and the statistics used to quantify the differences
between the waveforms.  We then calibrate the EOB model to the NR
waveforms in Sec.~\ref{sec:calibration}. In
Sec.~\ref{sec:firstorder-model}, we combine the results of
Sec.~\ref{sec:EOB-cal} with those of Ref.~\cite{Barausse:2011kb} to
build a prototype EOB model that interpolates between the calibrated
EOB waveforms and extends them to a larger region of the parameter
space. We also investigate how this prototype EOB model performs with
respect to two NR waveforms with nearly extremal spin, which were not
used in the calibration.  Finally, Sec.~\ref{sec:concl} summarizes our
main conclusions. In Appendix~\ref{sec:AppendixFactModes} we
explicitly write the factorized waveforms used in this work, including
spin effects.

\section{Effective-one-body dynamics and waveforms in the presence of
  spin effects}
\label{sec:EOB-model}

In this section, we define the spin EOB model that we will later
calibrate using NR waveforms. Henceforth, we use geometric units
$G=c=1$.

In the spin EOB model~\cite{Damour01c, Damour:024009, Barausse:2009xi,
  Nagar:2011fx, Barausse:2011ys} the dynamics of two black holes of
masses $m_1$ and $m_2$ and spins $\vS_1$ and $\vS_2$ is mapped into
the dynamics of an effective particle of mass $\mu =
m_1\,m_2/(m_1+m_2)$ and spin $\vS_*$ moving in a deformed Kerr metric
with mass $M =m_1+m_2$ and spin $\vS_\text{Kerr}$.  The position and
momentum vectors of the effective particle are described by $\vR$ and
$\vP$, respectively. Here, for convenience, we use the reduced
variables
\begin{equation}
  \vvr\equiv\frac{\vR}{M}\,, \qquad\qquad \vp\equiv\frac{\vP}{\mu}\,.
\end{equation}
Since we will restrict the discussion to spins aligned or anti-aligned
with the orbital angular momentum, we define the (dimensionless) spin
variables $\chi_i$ as $\vS_i\equiv\chi_i\,m_i^2\,\mathbf{\hat{L}}$,
where $\mathbf{\hat{L}}$ is the unit vector along the direction of the
orbital angular momentum.  We also write $\vS_\text{Kerr}\equiv
\chi_{\text{Kerr}} M^2 \mathbf{\hat{L}}$.

\subsection{The effective-one-body dynamics}
\label{sec:EOB-dyn}

In this paper we adopt the spin EOB Hamiltonian proposed in
Refs.~\cite{Barausse:2009aa, Barausse:2009xi, Barausse:2011ys}.  The
real (or EOB) Hamiltonian is related to the effective Hamiltonian
$H_{\text{eff}}$ through the relation
\begin{equation}
  \label{Hreal}
  H_{\text{real}}\equiv\mu\hat{H}_{\text{real}}=M\sqrt{1+2\nu\left(\frac{H_{\text{eff}}}{\mu}-1\right)}-M\,,
\end{equation}
where $H_{\text{eff}}$ describes the conservative dynamics of an
effective spinning particle of mass $\mu$ and spin $\vS^*$ moving in a
deformed Kerr spacetime of mass $M$ and spin $\vS_{\text{Kerr}}$.  The
symmetric mass ratio $\nu=\mu/M$ acts as the deformation parameter.
Through 3.5PN order in the spin-orbit coupling, the mapping between
the effective and real spin variables reads~\cite{Barausse:2009xi,
  Barausse:2011ys}
\begin{subequations}
  \begin{eqnarray}
    \label{spinmapping1}
    \vS_{\text{Kerr}} &=& \vS_1+\vS_2 \,, \\ 
    \label{spinmapping2}
    \vS^* &=& \frac{m_2}{m_1}\,\vS_1+\frac{m_1}{m_2}\,\vS_2 + \vDelta_{\sigma^*}^{(1)} +  \vDelta_{\sigma^*}^{(2)}\,,
  \end{eqnarray}
\end{subequations}
where $\vDelta_{\sigma^*}^{(1)}$ and $\vDelta_{\sigma^*}^{(2)}$ are
the 2.5PN and 3.5PN spin-orbit terms given explicitly in Eqs. (51) and
(52) of Ref.~\cite{Barausse:2011ys}. They depend on the dynamical
variables $\vvr$ and $\vp$, the spin variables $\vS_i$, and on several
gauge parameters. These parameters are present because of the large
class of canonical transformations that can map between the real and
effective descriptions.  Their physical effects would cancel out if
the PN dynamics were known at arbitrarily high orders; since this is
clearly not the case, the gauge parameters can have a noticeable
effect~\cite{Barausse:2011ys} and may in principle be used as spin EOB
adjustable parameters. In this paper however, we set all gauge
parameters to zero and introduce a spin EOB adjustable parameter at
4.5PN order in the spin-orbit sector by adding the following term to
Eq.~\eqref{spinmapping2}
\begin{equation}
  \vDelta_{\sigma^*}^{(3)}=\frac{d_{\text{SO}}\,\nu}{r^3}\,\left
    (\frac{m_2}{m_1}\,\vS_1+\frac{m_1}{m_2}\,\vS_2\right )\,.
\end{equation}
Here $d_{\text{SO}}$ is the spin-orbit EOB adjustable parameter. The
effective Hamiltonian reads~\cite{Barausse:2009xi}
\begin{equation}\label{Heff}
  \begin{split}
    \frac{H_{\text{eff}}}{\mu} &= \beta^i p_i + \alpha \sqrt{1 +
      \gamma^{ij} p_i p_j + \mathcal{Q}_4(\vp)} +
    \frac{H_{\text{SO}}}{\mu} + \frac{H_{\text{SS}}}{\mu} \\
    &\quad -\frac{1}{2Mr^5}(r^2\delta^{ij}-3r^i r^j)S_i^*S_j^* \,,
  \end{split}
\end{equation}
where the first two terms are the Hamiltonian of a nonspinning test
particle in the deformed Kerr spacetime, $\alpha$, $\beta^i$ and
$\gamma^{ij}$ are the lapse, shift and 3-dimensional metric of the
effective geometry and $\mathcal{Q}_4(\vp)$ is a non-geodesic term
quartic in the linear momentum introduced in
Ref.~\cite{Damour00a}. The quantities $H_{\text{SO}}$ and
$H_{\text{SS}}$ in Eq.~\eqref{Heff} contain respectively spin-orbit
and spin-spin couplings that are \textit{linear} in the effective
particle's spin $\boldsymbol{S}^*$, while the term
$-1/(2Mr^5)(r^2\delta^{ij}-3r^i r^j)S_i^*S_j^*$ is the leading-order
coupling of the particle's spin to itself, with $\delta^{ij}$ being
the Kronecker delta.  More explicitly, using
Ref.~\cite{Barausse:2009xi} we can obtain $H_{\text{SO}}$ and
$H_{\text{SS}}$ by inserting Eqs.~(5.31), (5.32),
Eqs.~(5.47a)--(5.47h), and Eqs.~(5.48)--(5.52) into Eqs.~(4.18) and
(4.19); $\alpha$, $\beta^i$ and $\gamma^{ij}$ are given by inserting
Eqs.~(5.36a)--(5.36e), Eqs.~(5.38)--(5.40) and Eqs.~(5.71)--(5.76)
into Eqs.~(5.44)--(5.46).  We will elucidate our choice of the quartic
term $\mathcal{Q}_4(\vp)$ at the end of this section, when introducing
the tortoise variables.

Following Ref.~\cite{Pan:2009wj}, we introduce another spin EOB
adjustable parameter in the spin-spin sector. Thus, we add to
Eq.~\eqref{Heff} the following 3PN term
\begin{equation}
  \frac{d_{\text{SS}}\,\nu}{r^4}\,\left
    (\frac{m_2}{m_1}\,\vS_1+\frac{m_1}{m_2}\,\vS_2\right )\cdot
  (\vS_1+\vS_2)\,,
\end{equation}
with $d_{\text{SS}}$ the spin-spin EOB adjustable parameter.  For what
concerns the nonspinning EOB sector, we adopt the following choice
for the EOB potentials $\Delta_t$ and $\Delta_r$ entering $\alpha$,
$\beta_i$ and $\gamma_{ij}$ (see Eq.~(5.36) in
Ref.~\cite{Barausse:2009xi}). The potential $\Delta_t$ is given
through 3PN order by
\begin{subequations}
  \begin{eqnarray}
    \label{deltatu}
    \Delta_t (u) &=& \frac{1}{u^2}\, \Delta_u(u)\,, \\
    \Delta_u(u) &=& A(u) + \chi^2_{\text{Kerr}}\,{u^2}\,,\\
    A(u) &=& 1 - 2 u + 2 \nu\, u^3 + \nu\,\left (\frac{94}{3} - \frac{41}{32} \pi^2\right)\, u^4\,,
    \label{deltauu}
  \end{eqnarray}
\end{subequations}
where $u \equiv 1/r$. Reference~\cite{Barausse:2009xi} suggested
rewriting the quantity $\Delta_u(u) $ as
\begin{eqnarray}
  \label{delta_t_1}
  \Delta_u(u) &=& \bar{\Delta}_u(u)\, \left [1 + \nu\,\Delta_0 + 
    \nu \,\log \left (1 + \Delta_1 \,u + \Delta_2\,u^2 \right. \right.
  \nonumber \\
  && \left. \left. + \Delta_3\,u^3 + \Delta_4\,u^4\right ) \right ]\,,
\end{eqnarray}
where $\Delta_i$ with $i = 1, 2, 3, 4$ are explicitly given in
Eqs.~(5.77)--(5.81) of Ref.~\cite{Barausse:2009xi}, and
\begin{subequations}
  \begin{align}
    \bar{\Delta}_u(u)=&\,\chi_{\text{Kerr}}^2\,\left(u -
      \frac{1}{r^{\text{EOB}}_{+}}\right)\,
    \left(u - \frac{1}{r^{\text{EOB}}_{-}}\right)\,,\\
    \label{eq:hor}
    r^{\text{EOB}}_{\pm} =&\, \left(1\pm
      \sqrt{1-\chi^2_\text{Kerr}}\right)\,(1-K\,\nu)\,.
  \end{align}
\end{subequations}
Here, $r^{\text{EOB}}_{\pm}$ are radii reducing to those of
the Kerr event and Cauchy horizons when the EOB adjustable parameter
$K$ goes to zero. The logarithm in Eq.~\eqref{delta_t_1} was
introduced in Ref.~\cite{Barausse:2009xi} to quench the divergence of
the powers of $u$ at small radii. Its presence also allows the
existence of an ISCO, a photon circular orbit (or light-ring), and a
maximum in the orbital frequency during the plunge. The reason for
modeling $\Delta_u(u)$ with Eq.~\eqref{delta_t_1} instead of using the
Pad\'e summation of $\Delta_u(u)$, as proposed in
Ref.~\cite{Damour:024009}, is threefold.  First, we did not want to
use the Pad\'e summation of $\Delta_u(u)$ because
Ref.~\cite{Pan:2009wj} found that for certain regions of the parameter
space spurious poles can appear. Second, although we could have
applied the Pad\'e summation only to $A(u)$ and used the Pad\'e
potential $A(u)$ calibrated to nonspinning waveforms in
Ref.~\cite{Pan:2011gk}, we want to take advantage of the good
properties of the potential \eqref{delta_t_1} during the late
inspiral, as found in Ref.~\cite{Barausse:2009xi}.  Third, we find it
useful to develop a variant of the EOB potential so that in the future
we can test how two different EOB potentials (both calibrated to NR
waveforms at high frequency) compare at low frequency.

Furthermore, for the potential $\Delta_r$ at 3PN order entering the
EOB metric components (5.36) in Ref.~\cite{Barausse:2009xi}, we choose
\begin{subequations}
  \begin{eqnarray}
    \label{eq:D}
    \Delta_r (u)&=& \Delta_t(u)\,D^{-1}(u)\,,\label{eq:deltaR}\\
    D^{-1}(u) &=& 1+\log[1 + 6 \nu\, u^2 + 2 (26 - 3 \nu)\, \nu\, u^3]\,.\nonumber \\
  \end{eqnarray}
\end{subequations}
Once expanded in PN orders, the EOB Hamiltonian \eqref{Hreal} with the
effective Hamiltonian defined in Eq.~\eqref{Heff} and the spin mapping
defined in Eqs.~\eqref{spinmapping1} and \eqref{spinmapping2},
reproduces all known PN orders---at 3PN, 3.5PN and 2PN order in the
nonspinning, spin-orbit and spin-spin sectors, respectively---except
for the spin-spin terms at 3PN and 4PN order, which have been recently
computed in Refs.~\cite{Porto:2006bt, Porto:2005ac, PR08a, PR08b,
  Porto:2010tr, Porto:2010zg, Levi:2008nh, Levi:2011eq}.  Furthermore,
in the test-particle limit the real Hamiltonian contains the correct
spin-orbit couplings linear in the test-particle spin, at \emph{all}
PN orders~\cite{Barausse:2009aa, Barausse:2009xi}.

Let $\hat{t}\equiv t/M$. In terms of the reduced Hamiltonian
$\hat{H}_{\text{real}}$, the EOB Hamilton equations are given in
dimensionless form by~\cite{Pan:2009wj}
\begin{subequations}
  \begin{eqnarray}
    \frac{d\vvr}{d\hat{t}}&=&\{\vvr,\hat{H}_{\text{real}}\}=\frac{\partial \hat{H}_{\text{real} }}{\partial \vp}\,,\\
    \frac{d\vp}{d\hat{t}}&=&\{\vp,\hat{H}_{\text{real}}\}+\hat{\bm{\mathcal{F}}}=-\frac{\partial \hat{H}_{\text{real}}}{\partial \vvr}
    +\hat{\bm{\mathcal{F}}}\,,
  \end{eqnarray}
\end{subequations}
where $\hat{\bm{\mathcal{F}}}$ denotes the non-conservative force that
accounts for radiation-reaction effects. Following
Ref.~\cite{Buonanno06}, we use~\footnote{The over-dot stands for
  $d/dt$.}
\begin{equation}
  \hat{\bm{\mathcal{F}}}=\frac{1}{\nu \hat{\Omega} |\vvr\times
    \vp|}\frac{dE}{dt}\vp\,,
\end{equation}
where $\hat{\Omega}\equiv M |\vvr\times\dot{\vvr}|/r^2$ is the
dimensionless orbital frequency and $dE/dt$ is the GW energy flux for
quasicircular orbits obtained by summing over the modes $(\ell,m)$ as
\begin{equation}\label{Edot}
  \frac{dE}{dt}=\frac{\hat{\Omega}^2}{8\pi}\sum_{\ell=2}^8\sum_{m=0}^{\ell}m^2\left|\frac{\mathcal{R}}{M}h_{\ell
      m}\right|^2\,.
\end{equation}
Here $\mathcal{R}$ is the distance to the source, and simply
eliminates the dominant behavior of $h_{\ell m}$. We sum over positive
$m$ modes only since $|h_{\ell,m}|=|h_{\ell,-m}|$. Expressions for the
modes $h_{\ell m}$ are given in the next section.  In this paper, we
restrict the calibration to non-precessing binaries, and thus we omit
the Hamilton equations of the spin variables.

It was demonstrated in previous work~\cite{Damour:2007cb, Damour2007}
that by replacing the radial component of the linear momentum $p_r
\equiv (\vp \cdot \vvr)/r$ with $p_{r^*}$, which is the conjugate
momentum of the EOB tortoise radial coordinate $r^*$, one can improve
the numerical stability of the EOB equations of motion. This happens
because $p_r$ diverges when approaching $r_{+}^{\text{EOB}}$ while
$p_{r^*}$ does not. In this paper we follow the definition of the EOB
tortoise radial coordinate in Appendix~A of
Ref.~\cite{Pan:2009wj}.\footnote{Note that all the formulas in
  Appendix~A of Ref.~\cite{Pan:2009wj} are written in physical
  dynamical variables, namely $\vR$ and $\vP$, while here we use
  reduced variables $\vvr$ and $\vp$.} However, when applying the
tortoise coordinate transformation to the quartic term in
Eq.~\eqref{Heff}, we get~\cite{Pan:2009wj}
\begin{equation}
  \label{Q4div}
  \mathcal{Q}_4(\vp) \propto \frac{p_{r^*}^4}{r^2}
  \frac{D^2}{\Delta_t^4} (r^2+\chi_{\text{Kerr}}^2)^4\,,
\end{equation}
which clearly diverges at $r=r^{\text{EOB}}_+$. As in the nonspinning
case~\cite{Damour:2007cb, Damour2007, Pan:2011gk}, we neglect
contributions higher than 3PN order and rewrite Eq.~\eqref{Q4div} as
\begin{equation}
  \mathcal{Q}_4(\vp) \propto \frac{p_{r^*}^4}{r^2}
  (r^2+\chi_{\text{Kerr}}^2)^4\,,
\end{equation}
which is well behaved throughout the EOB orbital evolution.

Lastly, we integrate the EOB Hamilton equations. In order to get rid
of any residual eccentricity when the EOB orbital frequency is close
to the initial frequency of the NR run, we start the EOB evolution at
large separation, say $50M$, and use the quasispherical initial
conditions developed in Ref.~\cite{Buonanno06}. We stop the
integration when the orbital frequency $\Omega$ reaches a maximum.

\subsection{The effective-one-body waveforms}
\label{sec:EOB-wave}

Following Refs.~\cite{Damour2007, Damour2009a, Buonanno:2009qa,
  Pan:2009wj, Pan:2011gk} we write the inspiral-plunge modes as
\begin{equation}
  h_{\ell m}^{\text{insp-plunge}}=h^{\text{F}}_{\ell m}\,N_{\ell m},
\end{equation}
where the $h_{\ell m}^{\text{F}}$ are the factorized modes developed
in Refs.~\cite{Damour2007, DIN, Pan2010hz}, and the $N_{\ell m}$ are
non-quasicircular (NQC) corrections that model deviations from
quasicircular motion, which is assumed when deriving the $h_{\ell
  m}^{\text{F}}$. The factorized modes read
\begin{equation}\label{hlm}
  h^\mathrm{F}_{\ell m}=h_{\ell m}^{(N,\epsilon)}\,\hat{S}_\text{
    eff}^{(\epsilon)}\, T_{\ell m}\, e^{i\delta_{\ell 
      m}}\left(\rho_{\ell m}\right)^\ell\,,
\end{equation}
where $\epsilon$ is the parity of the waveform. All the factors
entering the $h_{\ell m}^{\text{F}}$ can be explicitly found in
Appendix~\ref{sec:AppendixFactModes}. We emphasize here again that
despite the fact that the GW energy flux in the comparable-mass case
is known through 3PN order in the spin-orbit
sector~\cite{Blanchet:2011zv}, the spin-orbit couplings in the
factorized (or PN-expanded) modes have been computed only through
1.5PN order~\cite{Arun:2009, Pan2010hz}.  This limitation will degrade
the performances of our spin EOB model for prograde orbits and large
spin values, as already observed in the test-particle limit in
Refs.~\cite{Pan2010hz, Barausse:2011kb}.  To improve the knowledge of
spin effects in the GW modes, Refs.~\cite{Pan:2009wj, Pan2010hz} added
spin couplings in the test-particle limit through 4PN order in the
factorized waveforms.  However, since the mapping between the Kerr
spin parameter in the test-particle limit and the black-hole spins in
the comparable-mass case is not yet unambiguously
determined~\cite{Barausse:2009xi, Barausse:2011ys}, and since we do
not have many NR spinning waveforms at our disposal to test the
mapping, we decide not to include here the spinning
test-particle-limit couplings in the factorized waveforms computed in
Ref.~\cite{Pan2010hz}. We have checked before performing any
calibration that EOB models with or without test-particle spin effects
(with Kerr spin parameter $\chi_{\text{Kerr}}$) give similar
performances.

In all the calibrations of the nonspinning EOB model, two EOB
adjustable parameters were needed to calibrate the EOB Hamilton
equations---for example Refs.~\cite{Damour2009a, Pan:2011gk} used the
4PN and 5PN order coefficients in the EOB potential $A(r)$. As
discussed in the previous section, for the EOB model adopted in this
paper, the EOB nonspinning conservative dynamics depend so far only
on the adjustable parameter $K$. We introduce a second EOB adjustable
parameter in the non-conservative non-spinning EOB sector by adding a
4PN order non-spinning term in $\rho_{22}$ and denote the coefficient
of this unknown PN term by $ \rho_{22}^{(4)}$ [see
Eq.~\eqref{rho22}]. This adjustable parameter enters the EOB Hamilton
equations through the energy flux defined in Eq.~\eqref{Edot}.

As shown in Ref.~\cite{Pan:2011gk}, the NQC corrections of modes with
$(\ell,m) \neq (2,2)$ have marginal effects on the dynamics. Also, our
goal in this work is to calibrate only the $(2,2)$ mode, so in the
following we set $N_{\ell m}=1$ for $(\ell,m) \neq (2,2)$. We
have\footnote{Note that in Ref.~\cite{Barausse:2011kb} the $N_{\ell
    m}$ were written in terms of physical dynamical variables, rather
  than the reduced variables used here.}
%
%
\begin{equation}\label{NQC}
  \begin{split}
    N_{22} &= \Bigg[1 + \left( \frac{p_{r^*}}{r\,\hat{\Omega}}
    \right)^{\!2}\! \Bigg(a_1^{h_{22}}\! +\! \frac{a_2^{h_{22}}}{r}\! +\!
    \frac{a_3^{h_{22}}}{r^{3/2}} 
     \!+\!\frac{a_4^{h_{22}}}{r^{2}}\! +\! \frac{a_5^{h_{22}}}{r^{5/2}}
    \Bigg) \Bigg]\\
& \times \exp \Bigg[i \frac{p_{r^*}}{r\,\hat{\Omega}}
    \Bigg(b_1^{h_{22}} 
    + p_{r^*}^2 b_2^{h_{22}} \!+\! \frac{p_{r^*}^2}{r^{1/2}}
    b_3^{h_{22}}+ \frac{p_{r^*}^2}{r} b_4^{h_{22}} \Bigg) \Bigg],
  \end{split}
\end{equation}
where $a_i^{h_{22}}$ (with $i=1...5$) are the (real) NQC amplitude
coefficients and $b^{h_{22}}_i$ (with $i=1...4$) are the (real) NQC
phase coefficients. We will explain in detail how these coefficients
are determined at the end of this section.

The EOB merger-ringdown waveform is built as a linear superposition of
the quasinormal modes (QNMs) of the final Kerr black
hole~\cite{Buonanno00, Damour06, Buonanno-Cook-Pretorius:2007,
  Buonanno2007, DN2007b, DN2008, Buonanno:2009qa}, as
\begin{equation}\label{ringdown}
  h_{22}^{\text{merger-RD}}(t)=\sum_{n=0}^{N-1}
  A_{22n}\,e^{-i\sigma_{22n}(t-t_{\text{match}}^{22})}\,,
\end{equation}
where $N$ is the number of overtones, $A_{22n}$ is the complex
amplitude of the $n$-th overtone, and
$\sigma_{22n}=\omega_{22n}-i/\tau_{22n}$ is the complex frequency of
this overtone with positive (real) frequency $\omega_{22n}$ and decay
time $\tau_{22n}$. The complex QNM frequencies are known functions of
the mass and spin of the final Kerr black hole. Their numerical values
can be found in Ref.~\cite{Berti2006a}.  The mass and spin of the
final black hole, $M_f$ and $a_f$, can be computed through analytical
phenomenological formulas reproducing the NR predictions. Here, we
adopt the formulas given in Eq.~(8) of Ref.~\cite{Tichy2008} and in
Eqs.~(1) and (3) of Ref.~\cite{Barausse2009}.  We notice that the
formula for the final mass in Ref.~\cite{Tichy2008} was obtained using
numerical simulations of small-spin black-hole binaries with mildly
unequal masses. As a consequence, the formula is not very accurate for
the large-spin, unequal-mass binaries considered in this
paper. However, other formulas available in the literature are either
very accurate but only valid for equal-mass
binaries~\cite{Reisswig:2009vc}, or have not been yet extensively
tested against NR simulations~\cite{Kesden:2008ga, Lousto:2009mf}.
Thus, for the time being we stick with Eq.~(8) of
Ref.~\cite{Tichy2008}, but we plan to construct a better formula in
the future using all recent data in the literature.

Furthermore, we follow the hybrid matching procedure of
Ref.~\cite{Pan:2011gk} to fix the $N$ complex amplitude coefficients
$A_{22n}$ in Eq.~\eqref{ringdown}.  We set up $N$ complex linear
equations by imposing that the inspiral-plunge and merger-ringdown
waveforms $h_{22}^{\text{inspiral-plunge}}$ and
$h_{22}^{\text{merger-RD}}$ coincide on $N-2$ points (evenly sampled
over a range $[t_{\text{match}}^{22}-\Delta
t_{\text{match}}^{22},t_{\text{match}}^{22}]$) and that their time
derivatives $\dot{h}_{22}^{\text{inspiral-plunge}}$ and
$\dot{h}_{22}^{\text{merger-RD}}$ coincide at
$t_{\text{match}}^{22}-\Delta t_{\text{match}}^{22}$ and
$t_{\text{match}}^{22}$. As in previous works, we introduce the EOB
adjustable parameter $\Delta t_{\text{match}}^{22}$ which describes
the size of the comb over which we impose continuous and smooth
matching in order to determine the ringdown waveform.

In Refs.~~\cite{Buonanno:2009qa, Pan:2011gk, Barausse:2011kb}, pseudo
QNMs (pQNMs) were proposed and applied to moderate the rise of the EOB
GW frequency during the merger-ringdown transition---for example
Sec.~IIC of Ref.~\cite{Pan:2011gk} discussed in some detail the
advantage of using pQNMs for higher-order GW modes. In this paper, we
find it useful to introduce a pQNM for the $(2,2)$ mode. Therefore, we
choose $N\!=\!8$ in Eq.~\eqref{ringdown} and replace the highest
overtone in the summation with this pQNM.

Finally, we build the full inspiral-plunge-merger-ringdown EOB
waveform by joining the inspiral-plunge waveform
$h_{22}^{\text{inspiral-plunge}}(t)$ and the merger-ringdown waveform
$h_{22}^{\text{merger-RD}}(t)$ at the matching time
$t_{\text{match}}^{22}$ as
\begin{equation}
  \begin{split}
    h^{\text{EOB}}_{22}(t) &= h_{22}^{\text{inspiral-plunge}}(t)\,
    \theta(t_{\text{match}}^{22}-t) \\
    &\quad +h_{22}^{\text{merger-RD}}(t)\,
    \theta(t-t_{\text{match}}^{22}) \,.
  \end{split}
\end{equation}
\begin{figure}[!ht]
  \begin{center}
    \includegraphics*[width=0.45\textwidth]{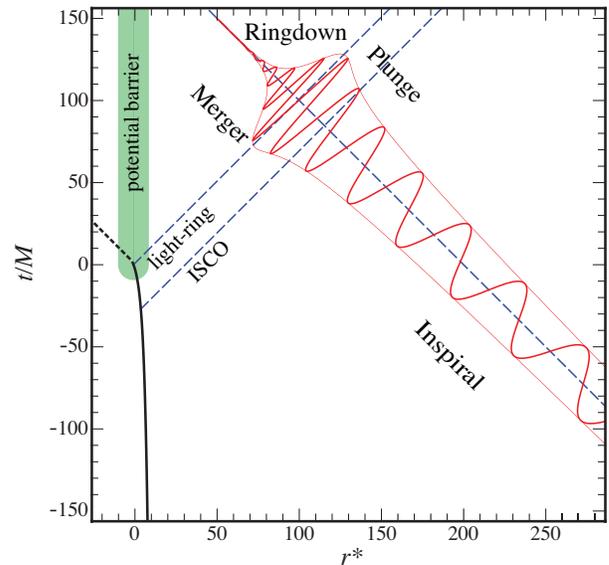}
    \caption{\label{fig:EOBSpacetime} We show in the spacetime diagram $(\hat{t},r^\ast)$ 
        the trajectory of the effective particle in the EOB description
(black solid line in the left part of the diagram)
 and the 
EOB (2,2) gravitational mode (red solid oscillating line) 
        for an equal-mass nonspinning black-hole binary. 
	Although we only need to evolve the EOB trajectory until the orbital frequency reaches its maximum (``light ring''),
	the model's dynamics allows the trajectory to continue to negative $r^\ast$ (short-dashed black line
in the left part of the diagram).
The blue dashed lines represent $\hat{t}\pm r^*=\textrm{const.}$ surfaces and ingoing/outgoing null rays. The EOB (2,2) mode is a function of the retarded time $\hat{t}-r^*$, plotted here orthogonal to $\hat{t}- r^*=\textrm{const.}$ surfaces, at a finite $\hat{t}+r^*$ distance. The two outgoing null rays are drawn at
        the $\hat{t}-r^*$ retarded times when the EOB particle 
        crosses the EOB ISCO and light-ring radii, respectively. The
        shaded green area is a rough sketch of the potential barrier around the newborn black hole.}
  \end{center}
\end{figure}

In Fig.~\ref{fig:EOBSpacetime}, we summarize how the inspiral-plunge--merger-ringdown EOB
  waveform is constructed. Beyond the ISCO, the quasi-circular inspiral waveform is followed by a
  short plunge waveform~\footnote{The number of gravitational-wave cycles during the 
    plunge scales roughly as $\nu^{-1/5}$~\cite{Buonanno00}.} where 
  the radial motion is no longer negligible and NQC corrections quickly become important. 
  The plunge ends roughly when the effective particle in the EOB description crosses the
  light-ring, which, in the nonspinning case, coincides approximately with 
  the peak of EOB orbital frequency $\hat{\Omega}$ and waveform amplitude $|h_{22}|$. 
  Until this moment, the GW radiation in the EOB description 
  is obtained directly from the motion of the effective particle.  
  After this moment that we identify as the merger, the direct emission of GWs 
  from the effective particle is strongly attenuated and filtered by the
  potential barrier formed around the newborn black hole. Thus, in the 
EOB description the merger-ringdown waveform is no longer obtained from the motion 
of the effective particle, but it is built through a superposition of QNMs. 
  This procedure of constructing the full EOB waveform, in particular replacing the direct 
emission with a superposition of QNMs beyond the light ring, was first proposed in Refs.~\cite{Buonanno00,
Buonanno06} for nonspinning and spinning comparable-mass black-hole binaries. 
It was inspired by the close limit approximation~\cite{price_pullin94} and results in Refs.~\cite{1971PhRvL..27.1466D,1972PhRvD...5.2932D} where it was observed that once the radially infalling particle 
is inside the potential barrier which peaks around the light ring, the 
direct gravitational radiation from the particle is strongly filtered by the potential barrier. 
Part of the energy produced in the strong merger-burst remains stored in the resonant 
cavity of the geometry, i.e., inside the potential barrier, and what is released outside is 
just the ringdown signal. The non-linear scattering of GW radiation (tails) against the
  curvature potential of the newborn black hole also contributes to the merger-ringdown 
  waveform. Currently, in the EOB description the merger-ringdown waveform is effectively 
  the tail of a $\delta$-function impulse at merger. When spin effects are present, the overall 
picture depicted in Fig.~\ref{fig:EOBSpacetime} survives, but with some differences due to the fact that the 
EOB light-ring position, peak of the orbital frequency $\hat{\Omega}$ and waveform amplitude $|h_{22}|$ 
can be displaced in time~\cite{Barausse:2011kb}. We notice that the physical picture of the merger-ringdown that emerged from the studies in 
Refs.~\cite{price_pullin94,1971PhRvL..27.1466D,1972PhRvD...5.2932D} and was incorporated in the EOB description in Refs.~\cite{Buonanno00,Buonanno06}, has also 
recently motivated the hybrid approach of Refs.~\cite{Nichols2010,Nichols:2011ih}.
 
\begin{table*}
  \caption{\label{tab:InputValues} Exact NR-input values used in the
    right-hand side of Eqs.~\eqref{NQCCond1}--\eqref{NQCCond5} to
    calibrate the EOB inspiral-plunge waveforms.}
  \begin{ruledtabular}
    \begin{tabular}{cccccccc}%
      $q$ & 1 & 1/2 & 1/3 & 1/4& 1/6 & 1 & 1 \\ %
      $\chi_1=\chi_2$ & 0 & 0& 0& 0& 0 & +0.43655 & -0.43757 \\[2pt] %
      \hline \\[-8pt]%
      $\; |h^{\text{NR}}_{22,\text{peak}}| \;$ & 0.3940 & 0.3446 & 0.2855
      & 0.2403 & 0.1810 & 0.3942 & 0.3935 \\[4pt] %
      $\; 10^4 M^2 \partial_t^2|h^{\text{NR}}_{22,\text{peak}}|\;$ & -10.3
      & -8.8 & -6.9 & -5.5 & -3.9 & -7.7 & -12.4 \\[4pt] %
      $\; M \omega^{\text{NR}}_{22,\text{peak}}\;$ & 0.3593 & 0.3467 &
      0.3324 & 0.3218 & 0.3084 & 0.3989 & 0.3342 \\[4pt] %
      $\; 10^3 M^2 \dot{\omega}^{\text{NR}}_{22,\text{peak}}\;$ & 11.3 &
      10.5 & 9.6 & 8.9 & 8.1 & 11.2 & 10.7 \\ %
    \end{tabular}
  \end{ruledtabular}
\end{table*}

We now continue our detailed review of how the EOB waveform is built and 
discuss  how we fix the NQC coefficients in Eq.~\eqref{NQC}.
Since we do not expect spin effects in the NQC correction until 1.5PN
order in either amplitude or phase, the coefficients $a^{h_{22}}_i$
with $i = 1,2$ and $b^{h_{22}}_i$ with $i=1,2$ only depend on $\nu$,
while $a^{h_{22}}_i$ with $i = 4,5$ and $b^{h_{22}}_i$ with $i = 3,4$
are functions of $\nu$ linearly proportional to the spins
$\chi_{1,2}$. The coefficient $a_3^{h_{22}}$ is given by the sum of a
nonspinning term (dependent only on $\nu$) and a spinning term
(proportional to the spins $\chi_{1,2}$). In Sec.~\ref{sec:EOB-ca} we
first calibrate the nonspinning waveforms, and then the spinning
ones.  Thus, we determine the ten coefficients in Eq.~\eqref{NQC} in
two steps. First, we set $\chi_1=\chi_2=0$, thus $a^{h_{22}}_i=0$
(with $i = 4,5$) and $b^{h_{22}}_i=0$ (with $i = 3,4$) and calculate
the values of the five NQC coefficients $a^{h_{22}}_i$ (with
$i=1,2,3$) and $b^{h_{22}}_i$ (with $i = 1,2$) by imposing the
following five conditions~\cite{Pan:2011gk, Barausse:2011kb}:
\begin{enumerate}
 \item Let $t_{\text{peak}}^{\Omega}$ be the time at which the EOB
  orbital frequency reaches its peak.  Then, the peak of the EOB
  $(2,2)$ mode must happen at the matching time $t_{\text{match}}^{22}
  = t_{\text{peak}}^{\Omega}+\Delta t^{22}_{\text{peak}}$, that is
  \begin{equation}
    \left. \frac{d |h^{\text{EOB}}_{22}|}{dt}\right
    |_{t_{\text{peak}}^{\Omega}+\Delta
      t^{22}_{\text{peak}}}=0 \label{NQCCond1}\,,
  \end{equation}
  where $\Delta t^{22}_{\text{peak}}$ is an EOB adjustable parameter,
  which will be specified in Sec.~\ref{sec:EOB-ca}. We note that in
  Ref.~\cite{Barausse:2011kb} the quantity $\Delta
  t^{22}_{\text{peak}}$ was computed by comparing the times at which
  the Teukolsky (2,2) mode and the EOB orbital frequency reach their
  peaks.  This was possible because the EOB trajectory was used in the
  Teukolsky equation to evolve the dynamics. However, in the NR
  simulation, we do not know what $\Delta t^{22}_{\text{peak}}$ is,
  because the EOB dynamics does not determine the NR dynamics.
 \item The amplitudes of the NR and EOB $(2,2)$ modes are the same,
  \begin{equation}
    |h^{\text{EOB}}_{22}(t_{\text{peak}}^{\Omega}+\Delta
    t_{\text{peak}}^{22})|=|h^{\text{NR}}_{22}(t_{\text{peak}}^{\text{NR}})|\label{NQCCond2}\,.
  \end{equation}

 \item The curvatures of the amplitudes of the NR and EOB $(2,2)$
  modes are the same,
  \begin{equation}
    \left. \frac{d^2 |h^{\text{EOB}}_{22}|}{dt^2}\right
    |_{t_{\text{peak}}^{\Omega}+\Delta
      t^{22}_{\text{peak}}}=\left. \frac{d^2
        |h^{\text{NR}}_{22}|}{dt^2}\right
    |_{t_{\text{peak}}^{\text{NR}}}\label{SpinCurvature} \,.
  \end{equation}

 \item The GW frequencies of the NR and EOB $(2,2)$ modes are the
  same,
  \begin{equation}
    \omega_{22}^{\text{EOB}}(t_{\text{peak}}^{\Omega}+\Delta
    t_{\text{peak}}^{22})=\omega_{22}^{\text{NR}}(t_{\text{peak}}^{\text{NR}})
    \,.
  \end{equation}

 \item The time derivatives of the GW frequency of the NR and EOB
  $(2,2)$ modes are the same,
  \begin{equation}
    \left. \frac{d \omega_{22}^{\text{EOB}}}{dt}\right
    |_{t_{\text{peak}}^{\Omega}+\Delta
      t_{\text{peak}}^{22}}=\left. \frac{d
        \omega_{22}^{\text{NR}}}{dt}\right
    |_{t_{\text{peak}}^{\text{NR}}}\label{NQCCond5} \,.
  \end{equation}
\end{enumerate}
We summarize in Table~\ref{tab:InputValues} all the NR-input values
that we use in the right-hand side of
Eqs.~\eqref{NQCCond2}--\eqref{NQCCond5}.  After the five nonspinning
NQC coefficients have been computed, we plug them back into the EOB
dynamics through the energy flux, start a new EOB evolution, generate
a new EOB $(2,2)$ mode, and calculate new NQC coefficients. We repeat
this procedure until the values of the NQC coefficients
converge. Then, when calibrating spinning waveforms, we set
$a^{h_{22}}_{i}$ and $b^{h_{22}}_{i}$ (with $i=1,2$), as well as the
nonspinning part of $a_3^{h_{22}}$, to the values just calculated for
$\chi_1=\chi_2=0$, and apply the five conditions above in an iterative
way, obtaining the final coefficients $a^{h_{22}}_{i}$ (with
$i=3,4,5$) and $b^{h_{22}}_{i}$ (with $i=3,4$). Note that in order to
generate GW templates, this procedure can be computationally
expensive, since to generate one EOB $(2,2)$ mode one has to evolve
the dynamics a few times. The current computational cost of generating
an EOB waveform long enough for the LIGO bandwidth varies between a
fraction of a second to a few seconds,\footnote{The time is measured
  by running a code that is not optimized in speed on a single CPU.}
depending on the masses. The iterative procedure can increase this
cost by a factor of a few.

In order for the NQC coefficients to be effective in correcting the
EOB mode peak, the latter has to occur in a region where the radial
motion is comparable to or at least $\roughly 30\%$ of the tangential
motion.  Such a condition is in principle not a necessary requirement
for the EOB model to work. In fact, the radial motion \textit{is}
expected to be strongly suppressed for almost extremal black holes, at
least in the test-particle limit, since the ISCO coincides with the
horizon for $\chi=1$~\cite{1972ApJ...178..347B}. However, if the
factorized (2,2) mode, given by Eq.~\eqref{hlm}, differs substantially
from the NR (2,2) mode because of the lack of high-order spin-orbit
terms, the inability of the NQC coefficient to change the waveform
during the plunge at high spins may prevent the EOB model to work
properly. This is because the NQC coefficients cannot artificially
compensate the missing higher-order spin orbit terms in the waveforms,
as they partially do at low spins. In fact, we will see that this
problem arises for $\chi_i \gtrsim 0.7$, making the EOB prototype
waveforms unreliable for large positive spins.

We list in Table~\ref{tab:adjparams} all the EOB adjustable parameters
that we exploit in this work to calibrate the EOB model to NR
simulations.

\section{Effective-one-body calibration}
\label{sec:EOB-ca}

In this section, we calibrate the EOB model using seven NR waveforms,
namely five nonspinning waveforms of mass ratios $q \equiv m_2/m_1 =
1,1/2,1/3,1/4$ and $1/6$ and two equal-mass spinning waveforms with
$\chi_1=\chi_2=+0.43655$ and $\chi_1=\chi_2=-0.43757$. The calibration
is achieved by minimizing the amplitude and phase differences between
the NR and EOB $(2,2)$ modes over the six EOB adjustable parameters:
$K$, $d_{\text{SO}}$ and $d_{\text{SS}}$ in the EOB conservative
dynamics, and $\rho_{22}^{(4)}$, $\Delta t^{22}_{\text{peak}}$,
$\Delta t_{\text{match}}^{22}$, $\omega_{22}^{\text{pQNM}}$ and
$\tau_{22}^{\text{pQNM}}$ in the EOB waveforms (see
Table~\ref{tab:adjparams}).

\subsection{Alignment of EOB and NR waveforms}
\label{sec:EOB-cal}

When calibrating NR and EOB waveforms, we first align the waveforms at
low frequency following the procedure of Refs.~\cite{Buonanno:2009qa,
  Pan:2009wj, Pan:2011gk}.  This procedure consists of minimizing the
square of the difference between the NR and EOB $(2,2)$-mode phases
$\phi^{\text{NR}}_{22}$ and $\phi^{\text{EOB}}_{22}$, integrated over
the time window $(t_1,t_2)$,
\begin{equation}\label{alignment}
  \int_{t_1}^{t_2} \left[
    \phi^{\text{EOB}}_{22}(t+t_0)+\phi_0-\phi^{\text{NR}}_{22}(t)
  \right]^2 dt\,,
\end{equation}
with respect to the time shift $t_0$ and phase shift $\phi_0$, where
it is understood that $\phi_{22}^{\text{EOB}}$ is computed for a
chosen set of adjustable parameters. The time window $(t_1,t_2)$
should: (i) begin as early as possible, where the NR and EOB GW-phase
evolutions agree best, (ii) begin late enough to avoid the junk
radiation present in the numerical simulation, (iii) be long enough to
average over numerical noise, and (iv) extend from peak to peak (or
trough to trough) over an integer number of oscillations in the GW
frequency, which are caused by the residual eccentricity in the
numerical initial conditions. In Table~\ref{tab:alignwindow}, we list
our choices of $(t_1,t_2)$ for the seven numerical waveforms at our
disposal. Each time window extends through 10 eccentricity oscillation
cycles in the numerical frequency evolution.

Let $\bar{\phi}_0$ and $\bar{t}_0$ be the alignment parameters. Then,
we define the phase and relative amplitude differences between the EOB
and NR (2,2) modes as follows:
\begin{equation}
  \Delta\phi(t)=\phi^{\text{EOB}}_{22}(t+\bar{t}_0)+\bar{\phi}_0-\phi^{\text{NR}}_{22}(t)\,,
\end{equation}
and
\begin{equation}
  \left(\frac{\Delta A}{A}\right)(t)=\frac
  {|h^{\text{EOB}}_{22}|(t+\bar{t}_0)}{|h^{\text{NR}}_{22}|(t)}-1\,.
\end{equation}
We then define the global phase and relative amplitude differences
over a time window $(t_1,t_3)$ with
\begin{equation}\label{phasediff}
  \Delta\phi_{\text{global}} =\max_{t\in (t_1,t_3)}|\Delta\phi(t)|\,,
\end{equation}
and
\begin{equation}\label{ampdiff}
  \left(\frac{\Delta A}{A}\right)_{\text{global}}=\max_{t\in
    (t_1,t_3)}\left| \left(\frac{\Delta A}{A}\right)(t)\right|\,.
\end{equation}
In the following, when measuring the difference between NR and EOB
inspiral-plunge waveforms we set $t_3=t_{\text{match}}^{22}$, while
when we measure the difference between full inspiral-merger-ringdown
waveforms we use $t_3=t_{\text{end}}$, where $t_{\text{end}}$ is
chosen as late as possible into the ringdown stage, but before
numerical errors due to gauge effects become
noticeable~\cite{Buonanno:2009qa}. We list the values of
$t_{\text{match}}^{22}$ and $t_{\text{end}}$ for the seven NR
waveforms in Table~\ref{tab:alignwindow}.

\begingroup
\begin{table}
  \caption{\label{tab:adjparams} Summary of adjustable parameters of
    the spin EOB model considered in this paper. The values of the
    EOB adjustable parameters used in this paper are given in
    Eqs.~\eqref{Delta22}, \eqref{combs}, \eqref{pQNM}, \eqref{A8},
    \eqref{K}, and \eqref{SpinParams}.  In addition, the NQC
    parameters $a_i^{h_{22}}$ and $b_i^{h_{22}}$ are fixed from
    NR-input values through
    Eqs.~\eqref{NQCCond1}--\eqref{NQCCond5}.}
  \begin{ruledtabular}
    \begin{tabular}{cc}
      EOB dynamics  & EOB waveform\\
      adjustable parameters & adjustable parameters\\ %
      \hline %
      $K$ & $\rho_{22}^{(4)}$ \\[3pt] %
      $d_{\text{SO}}, d_{\text{SS}}$ & $\Delta t_\text{match}^{22},
      \Delta t_{\text{peak}}^{22}$ \\[3pt] %
      & $\omega^\text{pQNM}_{22}, \tau^\text{pQNM}_{22}$ \\ %
    \end{tabular}
  \end{ruledtabular}
\end{table}
\endgroup

\subsection{Procedure to calibrate the EOB adjustable parameters}
\label{sec:calibration}

Recently, Ref.~\cite{Barausse:2011kb} computed the waveforms in the
small-mass-ratio limit by evolving a time-domain Teukolsky equation in
which the source term is evaluated using an EOB trajectory. It was
found that there exists a time difference between the Teukolsky
$(2,2)$-mode amplitude peak and the EOB orbital-frequency peak. This
difference is parametrized by the quantity $\Delta
t_{\text{peak}}^{22}$ introduced in Eq.~\eqref{NQCCond1}.  Table III
in Ref.~\cite{Barausse:2011kb} lists this difference as a function of
the Kerr spin parameter: for nonspinning and retrograde cases $-3M
\,\laq\, \Delta t^{22}_{\text{peak}}\,\laq\, 1.6 M$, while for
prograde cases $\Delta t^{22}_{\text{peak}}$ decreases quickly as
function of the spin. Let us consider $\chi_{\text{Kerr}}$, which
explicitly reads
\begin{equation}
  \chi_{\text{Kerr}} =
  (1-2\nu)\,\chi_{\text{S}}+\sqrt{1-4\nu}\,\chi_{\text{A}}\,,
\end{equation}
and also define
\begin{equation}
  \chi\equiv \chi_{\text{S}}+
  \chi_{\text{A}}\,\frac{\sqrt{1-4\nu}}{1-2\nu}\,, \label{chi}
\end{equation}
where $\chi_{\text{S,A}}\equiv(\chi_1\pm\chi_2)/2$. For an equal-mass,
equal-spin binary $(\nu=1/4, \chi_1=\chi, \chi_2=\chi)$ we have
$\chi_{\text{Kerr}}= \chi/2$, while in the test-particle limit we have
$\chi_{\text{Kerr}}= \chi$ (that is the spin parameter of the
background spacetime).  Therefore, inspired by the results in the
test-particle limit, we assume here that for an equal-mass, equal-spin
binary $\Delta t^{22}_{\text{peak}}$ depends on the black-hole spins
through $\chi$. Explicitly we choose
\begin{equation}
  \label{Delta22}
  \Delta t^{22}_{\text{peak}} =
  \begin{cases}
    -2.5M & \text{if } \chi \leq 0\,,\\
    -2.5M-1.77M \left(\frac{\chi}{0.437}\right)^4 & \text{if } \chi>0\,,\\
  \end{cases}
\end{equation}
which models qualitatively Table III in Ref.~\cite{Barausse:2011kb}.
Following Refs.~\cite{Buonanno:2009qa, Pan:2009wj, Pan:2011gk}, we
calibrate the EOB adjustable parameters in two steps. These steps are
performed for each of our seven calibration NR waveforms separately,
resulting in seven sets of calibration parameters.  First, for each of
the NR waveform at our disposal, we use $\Delta t_{\text{peak}}^{22}$
in Eq.~\eqref{Delta22}, insert the NR-input values from
Table~\ref{tab:InputValues} into
Eqs.~\eqref{NQCCond1}--\eqref{NQCCond5}, solve them iteratively for
the NQC coefficients, and calibrate $K$, $\rho_{22}^{(4)}$ (or
$d_{\text{SO}}$ and $d_{\text{SS}}$ if spins are present) by
minimizing Eq.~\eqref{phasediff} with
$t_3=t_{\text{match}}^{22}$. This process provides us with the EOB
inspiral-plunge waveform. Second, to obtain the EOB merger-ringdown
waveform, we calibrate the size of the comb $\Delta
t_{\text{match}}^{22}$ and the pQNM (complex) frequency by applying
Eq.~\eqref{phasediff} with $t_3=t_{\text{end}}$. As in
Ref.~\cite{Pan:2011gk}, we find that a constant value for the comb
size, notably
\begin{equation}
  \label{combs}
  \Delta t^{22}_{\text{match}} = 7.5M\,,
\end{equation}
gives a very good performance for all the different mass ratios and
spins. A detailed study of the pQNM (complex) frequency has revealed
that the best result is obtained when $\omega^{\text{pQNM}}_{22}$ lies
between the GW frequency $\omega^{\text{EOB}}_{22}M/M_f$ at
$t^{22}_{\text{match}}$ and the frequency of the least-damped QNM
$\omega_{220}$, and when $\tau^{\text{pQNM}}_{22}$ is (not much)
shorter than $\tau_{220}$. Specifically, we use the simple choice
\begin{subequations}
  \label{pQNM}
  \begin{eqnarray}
    \omega_{22}^{\text{pQNM}} &=&\frac{1}{2} \left[\omega^{\text{EOB}}_{22}(t_{\text{match}}^{22})\frac{M}{M_f} +\omega_{220}\right]\,,\\
    \tau_{22}^{\text{pQNM}} &=& \frac{3}{10} \tau_{220}\,,
  \end{eqnarray}
\end{subequations}
for all different mass ratios and spins. Before ending this section,
we discuss in more detail how we carry out the calibration of the
parameters $K$, $\rho_{22}^{(4)}$, for the nonspinning sector, and
the parameters $d_{\text{SO}}$, $d_{\text{SS}}$, for the spinning
sector.

\begin{table}
  \caption{\label{tab:alignwindow} We list the parameters $t_1$, $t_2$
    entering the alignment procedure defined in Eq.~\eqref{alignment},
    and the parameter $t_3$ (both $t_{\text{match}}^{22}$ and
    $t_{\text{end}}$) entering the computation of waveforms'
    differences in Eqs.~\eqref{phasediff} and \eqref{ampdiff}.}
  \begin{ruledtabular}
    \begin{tabular}{cccccccc}
      $q$ & 1 & 1/2 & 1/3 & 1/4& 1/6 & 1 & 1\\
      $\chi_1=\chi_2$ & 0 & 0& 0& 0& 0 & +0.43655 & -0.43757\\
      \hline
      $t_1/M \;$ & 820 & 770 & 570 & 670 & 870 & 800 & 610 \\
      $t_2/M \;$ & 2250 & 2255 & 1985 & 1985 & 2310 & 2150 & 1850 \\
      $t_{\text{match}}^{22}/M \;$ & 3943 & 3729 & 3515 & 3326 & 4892
      & 3367& 2402 \\
      $t_{\text{end}}/M \;$ & 3990 & 3770 & 3560 & 3370 & 4940 & 3410
      & 2430 \\
    \end{tabular}
  \end{ruledtabular}
\end{table}

\subsubsection{Calibrating nonspinning waveforms}
In general, the adjustable parameters $K$ and $\rho_{22}^{(4)}$ depend
on the mass ratio and we assume that they are polynomial functions of
$\nu$. In principle, we should determine $K(\nu)$ and
$\rho_{22}^{(4)}(\nu)$ by a global minimization of
$\Delta\phi_{\text{global}}$ and $(\Delta A/A)_{\text{global}}$ [as
defined in Eqs.~\eqref{phasediff} and \eqref{ampdiff} using $t_3 =
t_{\text{match}}^{22}$] with respect to the unknown coefficients
entering the $K(\nu)$ and $\rho_{22}^{(4)}(\nu)$ polynomials. However,
as in previous studies ~\cite{Damour2009a, Pan:2011gk}, we find a
strong degeneracy among the EOB adjustable parameters, when
calibrating each mass ratio separately. The degeneracy is partially
broken when we combine all the available mass ratios together, but it
is not completely lifted. In particular, different choices of $K(\nu)$
and $\rho_{22}^{(4)}(\nu)$ lead to EOB models that can match equally
well with NR waveforms. We are thus relieved from a rigorous yet
expensive global search and follow a simplified procedure to find
satisfactory $K(\nu)$ and $\rho_{22}^{(4)}(\nu)$.  First, we locate
two points $(0.8154,-35)$ and $(1.188,-20)$ in the
$K$--$\rho_{22}^{(4)}$ plane where $\Delta\phi_{\text{global}}<0.1$
rad and $(\Delta A/A)_{\text{global}}<0.1$ for $q=1$ and $q=1/6$
($\nu=0.25$ and $\nu=0.1224$), respectively. We then determine a
linear function $\rho_{22}^{(4)}(\nu)$ by imposing that
$\rho_{22}^{(4)}(0.25)=-35$ and $\rho_{22}^{(4)}(0.1224)=-20$, leading
to
\begin{equation}
  \rho_{22}^{(4)}(\nu) = -5.6 - 117.6\,\nu\,.\label{A8}
\end{equation}
At $q=1/2,1/3$ and $1/4$, we choose $\rho_{22}^{(4)}$ according to
Eq.~\eqref{A8} and determine the value of $K$ that minimizes
$\Delta\phi_{\text{global}}$ and a range of $K$ values that satisfy
$\Delta\phi_{\text{global}}<0.1$ rad.

\begin{figure}[!ht]
  \begin{center}
    \includegraphics*[width=0.45\textwidth]{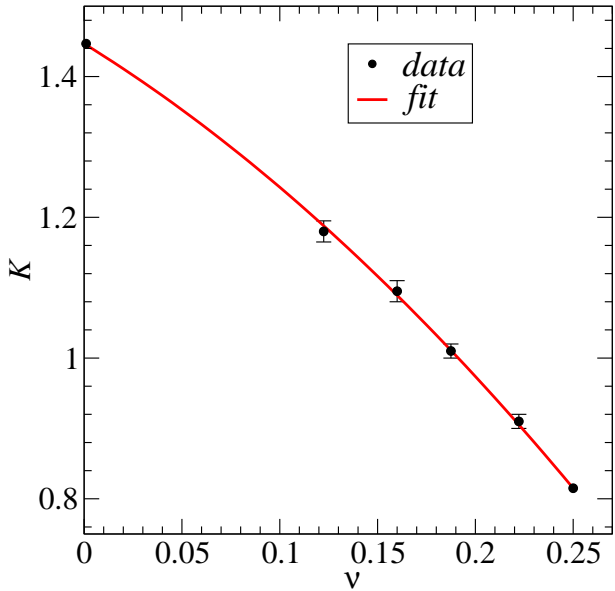}
    \caption{ \label{fig:KFit} We show the quadratic fit in $\nu$ for
      the adjustable parameter $K$. This parameter is calibrated using
      the five nonspinning NR waveforms, assuming
      $\rho_{22}^{(4)}(\nu)$ in Eq.~\eqref{A8}. The error bars are
      determined by the intersection of the contours of
      $\Delta\phi_{\text{global}} = 0.1$ rads with
      $\rho_{22}^{(4)}(\nu)$ for each mass ratio considered.}
  \end{center}
\end{figure}

We now have a complete set of calibration parameters for each of our
nonspinning NR waveforms.  In order to obtain calibration parameters
that interpolate between the NR waveforms, we build a least-squares
fit quadratic in $\nu$ against these $K$ values. By construction, we
fix two of the three free parameters in the fit by requiring that in
the test-particle limit $K(\nu)$ reproduces the ISCO shift of
Refs.~\cite{BarackSago09, Barausse:2009xi, LeTiec:2011dp} and that the
optimal equal-mass value $K(0.25)$ is recovered exactly. Even with
these two constraints and just one free parameter to fit, the
residuals are within $1\%$ (see Fig.~\ref{fig:KFit}). We find
\begin{equation}
  K(\nu) = 1.447 - 1.715\,\nu - 3.246\,\nu^2\,.\label{K}
\end{equation}
Finally, since the iterative procedure to compute the NQC coefficients
through Eqs.~\eqref{NQCCond1}--\eqref{NQCCond5} can be expensive, we
have parametrized them through quadratic fits, finding rather small
residuals. Explicitly, we obtain
\begin{subequations}
  \label{NQCNS}
  \begin{eqnarray}
    a_1^{h_{22}} &=& -12.68 + 75.42\,\nu - 106.6\,\nu^2,\\\label{a1NS}
    a_2^{h_{22}} &=& 101.5 - 757.3\,\nu + 1473\,\nu^2,\\
    a_3^{h_{22}} &=& -107.7 + 857.6\,\nu - 1776\,\nu^2,\\
    b_1^{h_{22}} &=& -1.464 + 12.82\,\nu - 60.10\,\nu^2,\\
    b_2^{h_{22}} &=& 7.477 - 85.26\,\nu + 353.3\,\nu^2. \label{b2NS}
  \end{eqnarray}
\end{subequations}
\subsubsection{Calibrating spinning waveforms}

When calibrating the EOB inspiral-plunge waveforms to the two NR
equal-mass, equal-spin waveforms at our disposal
($\chi_1=\chi_2=+0.43655$ and $\chi_1=\chi_2=-0.43757$), we use the
nonspinning EOB adjustable parameters $K$ and $\rho_{22}^{(4)}$ in
Eqs.~\eqref{K}-\eqref{A8}, and calibrate the spinning EOB adjustable
parameters $d_{\text{SO}}$ and $d_{\text{SS}}$. We reach this goal by
building contour plots in the plane $d_{\text{SO}}$--$d_{\text{SS}}$
for $\Delta \phi_{\text{global}}$ in Eq.~\eqref{phasediff} with $t_3 =
t_{\text{match}}^{22}$.  We find that the contours of $\Delta
\phi_{\text{global}} = 0.2$ rads associated with the two NR spinning
waveforms intersect each other for the following choice of the
adjustable parameters
\begin{equation}
  d_{\text{SO}}=-69.5\,, \quad \quad
  d_{\text{SS}}=2.75\,.\label{SpinParams}
\end{equation}
Note that when computing the spinning NQC coefficients, we use the NQC
coefficients parametrized in Eq.~\eqref{NQCNS}, and solve iteratively
the five conditions \eqref{NQCCond1}--\eqref{NQCCond5} for
$a_i^{h_{22}}$ ($i=3,4,5$) and $b_i^{h_{22}}$ ($i=3,4$).\footnote{Note
  that the NQC coefficient $a_3^{h_{22}}$ is solved for twice, first
  in the nonspinning calibration and then in the spinning one.}

\section{A prototype effective-one-body model for non-precessing
  spinning waveforms}
\label{sec:firstorder-model}

We now build on the results of Sec.~\ref{sec:EOB-ca}, and also on
recent outcomes of small-mass-ratio simulations produced by the
Teukolsky equation~\cite{Barausse:2011kb}, to construct a
self-contained set of prescriptions to generate EOB
inspiral-merger-ringdown waveforms in a larger region of the parameter
space $(\nu,\chi_1,\chi_2)$ of the binary.

\begingroup
\setlength{\tabcolsep}{5pt}
\begin{table*}
  \begin{minipage}{0.7\linewidth}
    \caption{\label{tab:InputValuesFits} Fits of the NR-input values
      $f^{\text{NR}}$ that are used to build the global fits in
      Eq.~\eqref{f} for the test-particle and equal-mass limits.}
    \begin{ruledtabular}
      \begin{tabular}{ccc}
        $f^{\text{NR}}$ & Curve & Fit \\
        \hline \\[-6pt]
        \multirow{2}{*}{$|h_{22,\text{peak}}^{\text{NR}}|$} &
        $(\nu=0,\chi)$ & 0 \\ 
        & $(\nu=1/4,\chi)$ & 0.3961\\[6pt]
        \multirow{2}{*}{$M^2 \partial_t^2|h_{22,\text{peak}}^{\text{NR}}|$}
        & $(\nu=0,\chi)$ & 0 \\[2pt]
        & $(\nu=1/4,\chi)$ & $10^{-3}\times (-1.007 + 0.5415 \chi)$
        \\[6pt]
        \multirow{2}{*}{$M \omega_{22,\text{peak}}^{\text{NR}}$} &
        $(\nu=0,\chi)$  & $0.2758 - 0.08898\log(1-\chi)$ \\[2pt]
        & $(\nu=1/4,\chi)$ &  $0.3604 + 0.08242 \chi + 0.02794 \chi^2$
        \\[6pt]
        \multirow{2}{*}{$M^2 \dot{\omega}_{22,\text{peak}}^{\text{NR}}$} &
        $(\nu=0,\chi)$  & $10^{-3}\times [5.953+(0.7199+1.210 \chi)
        \log(1 - \chi)]$ \\[2pt]
        & $(\nu=1/4,\chi)$ & 0.01113 \\[3pt]
      \end{tabular}
    \end{ruledtabular}
  \end{minipage}
\end{table*}
\endgroup

\subsection{Interpolating the EOB model outside the domain of
  calibration}
\label{sec:interpolation}

Since we only have seven NR waveforms at our disposal (and just two of
them with spins), when extending the EOB model to regions of the
parameter space without NR waveforms, we are forced to make
assumptions on the behavior of the adjustable parameter $\Delta
t_{\text{peak}}^{22}$ and the NR-input values in
Table~\ref{tab:InputValues}. In this work we assume that the 3
dimensional space $(\nu,\chi_1,\chi_2)$ can be treated as the 2
dimensional space $(\nu,\chi)$.  [Note that $\nu \in [0,1/4]$ and
$\chi \in [-1,1]$.] More specifically, given a binary described by the
parameters $(\nu,\chi_1,\chi_2)$ having in general $\chi_1 \neq
\chi_2$, we consider an auxiliary equal-spin binary with parameters
$(\nu,\chi,\chi)$, where $\chi$ is defined as in Eq.~\eqref{chi}.
With this choice, the auxiliary binary has the same value of
$\chi_{\text{Kerr}}$ as the original binary. We stress that the
auxiliary binary is used only to extend the EOB adjustable parameters
and the NR-input values to regions of the parameter space in which we
do not have NR results. Of course the EOB dynamics and waveforms are
computed for the original binary, not the auxiliary one.

Thus, in the prototype EOB model, the EOB adjustable parameter $\Delta
t_{\text{peak}}^{22}$ in Eq.~\eqref{Delta22} is evaluated using for
$\chi$ the value from Eq.~\eqref{chi}. To compute the spinning NQC
coefficients in the prototype model, we need to prescribe the input
values in the right-hand side of
Eqs.~\eqref{NQCCond2}--\eqref{NQCCond5} using the parameters of the
auxiliary binary. We proceed as follows. We only have knowledge of the
NR-input values at merger for a few regions of the $(\nu, \chi)$
parameter space. We can obtain the NR-input values along the curve
$(\nu=0,\chi)$ from the Teukolsky waveforms of
Ref.~\cite{Barausse:2011kb}. In particular, both
$|h_{22,\text{peak}}^{\text{NR}}|$ and
$\partial_t^2|h_{22,\text{peak}}^{\text{NR}}|$ are set to 0 (since
they are proportional to $\nu$), while for
$\omega_{22,\text{peak}}^{\text{NR}}$ and
$\dot{\omega}_{22,\text{peak}}^{\text{NR}}$ we use the data in Table V
of Ref.~\cite{Barausse:2011kb}.  We can extract the peak information
along the curve $(\nu=1/4,\chi)$ from the three equal-mass waveforms
used in the calibration of this paper, together with the two nearly
extremal spin cases $\chi_1=\chi_2=-0.94905$ and
$\chi_1=\chi_2=+0.9695$ (not used for the calibration of the
adjustable parameters $d_{\text{SO}}$ and $d_{\text{SS}}$), which we
will discuss in Sec.~\ref{sec:perf-otherwaveforms}. Along the curve
$(\nu,\chi=0)$ we can use the NR-input values of the nonspinning
waveforms from Refs.~\cite{Barausse:2011kb, Pan:2011gk}.  In
Table~\ref{tab:InputValuesFits} we list the fits for each NR-input
value
$f^{\text{NR}}\in\{|h_{22,\text{peak}}^{\text{NR}}|, \partial_t^2|h_{22,\text{peak}}^{\text{NR}}|,
\omega_{22,\text{peak}}^{\text{NR}},
\dot{\omega}_{22,\text{peak}}^{\text{NR}}\}$ in the test-particle and
equal mass limits.  Along the nonspinning profile, fits quadratic in
$\nu$ give a good description of the exact NR-input values, hence we
assume that the dependence of $f^{\text{NR}}$ on $\nu$ is quadratic as
well and has the simple form
\begin{equation}
  f^{\text{NR}}(\nu,\chi)=c_2(\chi)\,\nu^2+c_1(\chi)\,\nu+c_0(\chi)\,.
\end{equation}
We can fix two of the coefficients $c_i$ by imposing that the
test-particle limit and equal-mass cases are exactly recovered when
$\nu=0$ and $\nu=1/4$, respectively. We can fit the third coefficient
to the exact NR-input values along the nonspinning direction. This
means that the fits along the nonspinning profile are not exactly
recovered by the global fits $f^{\text{NR}}(\nu,\chi)$, but we find
that the residuals are negligible. Explicitly, we fit $c_1$ in the
following expression
\begin{eqnarray}
  f^{\text{NR}}(\nu,0;c_1)&=&\{16[f^{\text{NR}}(1/4,0)-f^{\text{NR}}(0,0)]-4c_1\}\,\nu^2\nonumber\\
  &+&c_1 \nu+f^{\text{NR}}(0,0)\,,
\end{eqnarray}
and denote the fitted value with $\bar{c}_1$. Finally, we extend the
result outside the nonspinning profile assuming that the global fit
reads
\begin{eqnarray}
  f^{\text{NR}}(\nu,\chi)&=&\{16[f^{\text{NR}}(1/4,\chi)-f^{\text{NR}}(0,\chi)]-4\bar{c}_1\}\,\nu^2\nonumber\\
  &+&\bar{c}_1 \nu+f^{\text{NR}}(0,\chi)\,.\label{f}
\end{eqnarray}
In Table~\ref{tab:c1bar} we list the values of $\bar{c}_1$ for the
four NR-input values that are needed to compute the right-hand sides
in Eqs.~\eqref{NQCCond2}--\eqref{NQCCond5}.
\begin{table}
  \caption{\label{tab:c1bar} Fitted values of $\bar{c}_1$ for the four
    NR-input values as defined in Eq.~\eqref{f}.}
  \begin{ruledtabular}
    \begin{tabular}{ccccc}
      &  $|h_{22,\text{peak}}^{\text{NR}}|$ &
      $M^2 \partial_t^2|h_{22,\text{peak}}^{\text{NR}}|$ & $M
      \omega_{22,\text{peak}}^{\text{NR}}$ & $M^2
      \dot{\omega}_{22,\text{peak}}^{\text{NR}}$ \\[2pt]
      \hline
      $\; \bar{c}_1 \;$ & 1.355 & $-2.5 \times 10^{-3}$ & 0.1935 &
      0.01204 \\
    \end{tabular}
  \end{ruledtabular}
\end{table}

Having in hand $\Delta t_{\text{peak}}^{22}$ and the NR-input values,
we complete the construction of the prototype EOB model by fixing the
EOB adjustable parameters $K$, $\rho_{22}^{(4)}$, and $d_{\text{SO}}$,
$d_{\text{SS}}$ to the values in Eqs.~\eqref{K}-\eqref{A8} and
\eqref{SpinParams}, respectively, employing the pQNM (complex)
frequency in Eq.~\eqref{pQNM}, the comb size in Eq.~\eqref{combs}, and
the NQC coefficients in Eqs.~\eqref{NQCNS}.

To test the robustness of the construction of the quantity
$f^{\text{NR}}(\nu,\chi)$, we study how the spinning NQC coefficients
change across the plane $(\nu,\chi)$. We focus on binaries with
$\chi_1=\chi_2=\chi$.  We compute iteratively the NQC amplitude
coefficients $a_i^{h_{22}}$ (with $i=3,4,5$) for different mass ratios
in the range $1/100 \leq q\leq 1$ and for different spins in the range
$-1 \leq \chi_i \lesssim 0.7$ ($i=1,2$). Typically, we get convergence
of the NQC coefficients within five iterations. Unfortunately, we
cannot span larger, positive values of $\chi_i$ since the NQC
corrections tend to diverge as the spin magnitude grows in the
prograde case. The reason is that they become less effective in
reshaping the EOB (2,2) peak as prescribed by the fits
$f^{\text{NR}}(\nu,\chi)$. This happens because the peak of the EOB
(2,2) mode occurs too early in the evolution when the orbital motion
is still quasicircular.  Hence the NQC coefficients must be very large
to compensate for the small values of $p_{r^*}/(r \hat{\Omega})$ and
be able to reshape the EOB (2,2) amplitude around the peak in a
satisfactory way. As discussed earlier, this would not be a problem in
principle if higher-order spin-orbit terms were known in the
factorized waveforms, but, as a result of the lack of knowledge of
those, our EOB prototype waveforms are reliable only up to $\chi_i
\lesssim 0.7$.

\subsection{Performance for nonspinning waveforms}
\begin{figure}[!ht]
  \begin{center}
    \includegraphics*[width=0.45\textwidth]{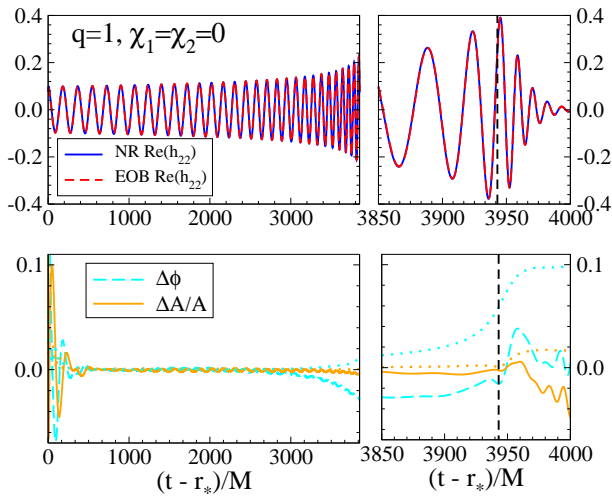}
    \caption{\label{fig:NS_q_1} Comparison of the NR and EOB (2, 2)
      mode for $q\!=\!1$, $\chi_1\!=\!\chi_2\!=\!0$.  In the upper
      panels we show the comparison between the real part of the two
      waveforms, zooming into the merger region in the upper right
      plot. In the lower panels we show the dephasing and relative
      amplitude difference over the same time ranges as the upper
      panels. A vertical dashed line marks the position of the NR
      amplitude peak. The dotted curves are the NR errors.}
  \end{center}
\end{figure}
\begin{figure}[!ht]
  \begin{center}
    \includegraphics*[width=0.45\textwidth]{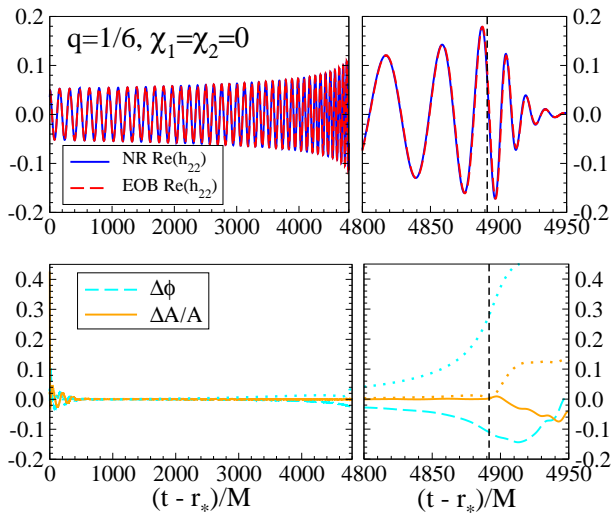}
    \caption{\label{fig:NS_q_6} Same as in Fig.~\ref{fig:NS_q_1} but
      for $q=1/6$, $\chi_1=\chi_2=0$.}
  \end{center}
\end{figure}
In Figs.~\ref{fig:NS_q_1} and \ref{fig:NS_q_6} we show how the
inspiral-merger-ringdown EOB waveforms computed according to the
prescriptions of Sec.~\ref{sec:interpolation} compare with the NR
waveforms for two representative mass ratios $q=1,1/6$. In general,
for all the nonspinning waveforms we find that the dephasing is
typically within $0.1$ rads up until $t_{\text{match}}^{22}$ (merger
time) and always within $0.2$ rads when including the ringdown stage.
The figures also show in dotted lines the NR phase and amplitude
errors obtained by combining the extrapolation and resolution errors
in quadrature. We notice that the EOB and NR amplitudes' agreement is
remarkably good up to the merger time, while during the ringdown the
relative amplitude difference may grow up to about $15\%$, approaching
the estimated NR error.

In Ref.~\cite{Pan:2011gk} the authors calibrated a different version
of the nonspinning EOB model to the same set of nonspinning NR
waveforms used in this paper, the main difference between the two EOB
models being the choice of the EOB potential $A(r)$, as we discussed
in Sec.~\ref{sec:EOB-dyn}. We find that the difference between the EOB
inspiral-merger-ringdown waveforms and the NR waveforms in
Ref.~\cite{Pan:2011gk} is comparable to and for some mass ratios
marginally worse than what we have achieved in this work using the
prototype EOB model. The only noticeable qualitative difference is
that the phase error of the prototype EOB model accumulates more
slowly during the merger-ringdown transition because of the
introduction of the pQNM in the $(2,2)$ mode. We point out that the
inclusion of the pQNM (complex) frequency in the EOB merger-ringdown
waveform is not strictly needed for the nonspinning case, but we use
it even in this case for uniformity with the spinning sector, where
the pQNM frequency is instead crucial.

We can quantify the differences between NR and EOB waveforms by
computing the mismatch ($\MM$), as defined in Eq.~(43) of
Ref.~\cite{Pan:2011gk}, which is one minus the overlap between two
waveforms, weighted by the noise spectral density of the detector and
maximized over the initial time, phase and binary parameters.  If we
use an Advanced LIGO noise curve, named \texttt{ZERO\_DET\_HIGH\_P} in
Ref.~\cite{Shoemaker2009}, we obtain that the $\MM$, maximizing only over
the initial phase and time, is always smaller than 0.001 when the
binary total mass varies between $20 M_\odot$ and $200 M_\odot$. For
these total masses, the NR waveforms start in band. We taper them
using the Planck-taper window function~\cite{McKechan:2010kp} to
reduce numerical artifacts. The width of the window function is set to
the length of NR waveforms, ranging from $0.35(M/20M_\odot)$ to
$0.65(M/20M_\odot)$ seconds. The window function smoothly rises from 0
to 1 in the first 0.0625 seconds and falls from 1 to 0 in the last
0.0125 seconds. We restrict the $\MM$ integration to the frequency band
for which NR waveform is available.

\subsection{Performance for spinning waveforms}

\begin{figure}[!ht]
  \begin{center}
    \includegraphics*[width=0.45\textwidth]{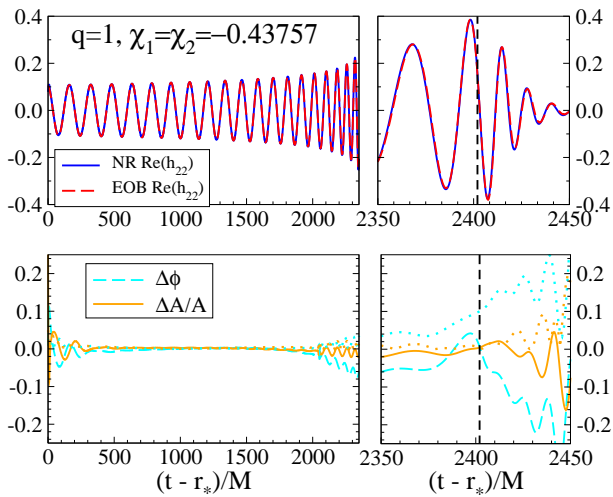}
    \caption{\label{fig:DD} Same as in Fig.~\ref{fig:NS_q_1} but for
      $q=1$, $\chi_1=\chi_2=-0.43655$.}
  \end{center}
\end{figure}

\begin{figure}[!ht]
  \begin{center}
    \includegraphics*[width=0.45\textwidth]{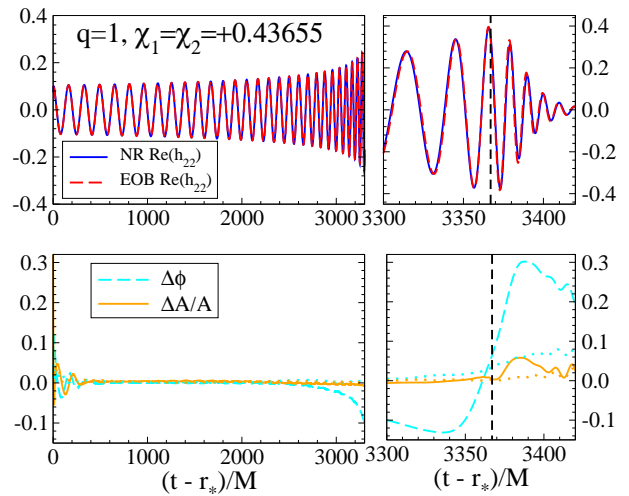}
    \caption{\label{fig:UU} Same as in Fig.~\ref{fig:NS_q_1} but for
      $q=1$, $\chi_1=\chi_2=+0.43756$.}
  \end{center}
\end{figure}

In Figs.~\ref{fig:DD} and \ref{fig:UU} we present the results of the
prototype EOB model for the two moderately spinning waveforms at our disposal. We
observe that the choice \eqref{SpinParams} gives a larger dephasing
for $\chi_1=\chi_2=+0.43655$ than for $\chi_1=\chi_2=-0.43757$ or the
nonspinning runs. In fact at the merger time the dephasing for the
$\chi_1=\chi_2=+0.43655$ waveform grows beyond the NR error.  For the
amplitude, we instead get a similar performance, on the same level as
the other runs.  The worse performance of the $\chi_1=\chi_2=+0.43655$
waveform can be explained by the more relativistic nature of this
run. In fact, in this case the EOB ISCO moves to smaller radial
separations as the spin parameter $\chi$ increases towards positive
values (aligned runs). On the other hand, for negative values of
$\chi$ (anti-aligned runs) the EOB ISCO moves outwards to a less
relativistic regime and one expects a better behavior of the EOB
model. This expectation is confirmed by the calibration of the
$\chi_1=\chi_2=-0.43757$ run, for which we find that very good
performances can be achieved in large regions of the EOB adjustable
parameter space.  Fig.~\ref{fig:DD} shows that in this case the
dephasing is well within the NR error at the merger time.
For these spinning waveforms, we obtain that the $\MM$, maximizing only
over the initial phase and time, is always smaller than 0.003 when the
binary total mass varies between $20 M_\odot$ and $200 M_\odot$.

\subsection{Performance for nearly extremal spin waveforms}
\label{sec:perf-otherwaveforms}

Here we compare the EOB waveforms of the prototype model developed in
Sec.~\ref{sec:interpolation}, against two equal-mass NR waveforms with
nearly extremal spins: $\chi_1=\chi_2=-0.94905$ and
$\chi_1=\chi_2=+0.9695$~\cite{Lovelace2010, Lovelace:2011nu}.  We
stress that these NR waveforms were not used when calibrating the spin
EOB adjustable parameters $d_{\text{SO}}$ and $d_{\text{SS}}$ in
Eq.~\eqref{SpinParams}. The only information that we used from these
two nearly extremal spin waveforms was their NR-input values when
building the fits $f^{\text{NR}}(\nu,\chi)$.

\begin{figure}[!ht]
  \begin{center}
    \includegraphics*[width=0.45\textwidth]{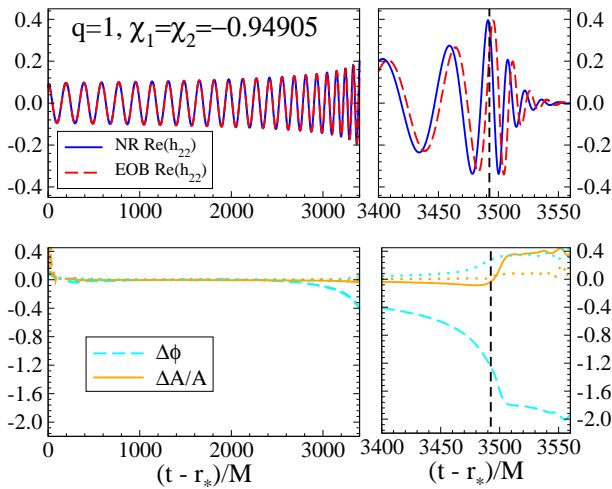}
    \caption{ \label{fig:ExtDD} Same as in Fig.~\ref{fig:NS_q_1} but
      for $q=1$, $\chi_1=\chi_2=-0.94905$. This NR waveform was
      \emph{not} used to calibrate the adjustable parameters
      $d_{\text{SO}}$ and $d_{\text{SS}}$.  Alignment between the NR
      and EOB waveforms was performed using Eq.~\eqref{alignment},
      with $t_{1} = 860\,M$ and $t_{2} = 2470\, M$.}
  \end{center}
\end{figure}

As already discussed, when the spins are anti-aligned, the EOB ISCO
moves towards larger radial separations, so that the binary is less
relativistic throughout its orbital evolution as compared to the
aligned configurations.  Therefore, we expect that in this case the
EOB model is more effective. The results in Fig.~\ref{fig:ExtDD} for
the case $\chi_1=\chi_2=-0.94905$ confirm this expectation. The
dephasing grows up to about 2 rads during the ringdown, while the
relative amplitude difference grows up to about $40\%$.  Despite the
large phase difference at merger, we find that, even without
maximizing over the binary parameters but only the initial phase and
time, the $\MM$ is always smaller than $0.005$ for systems with total
mass between 20$M_{\odot}$ and 200$M_{\odot}$.

\begin{figure}[!ht]
  \begin{center}
    \includegraphics*[width=0.45\textwidth]{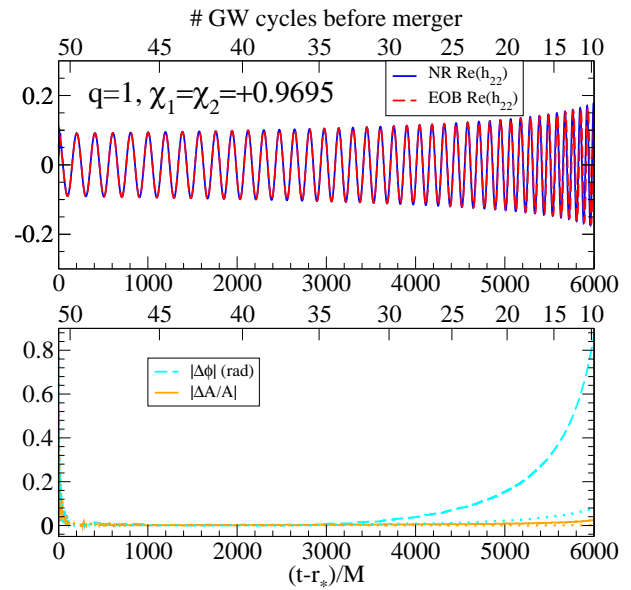}
    \caption{ \label{fig:ExtUU} Same as in Fig.~\ref{fig:NS_q_1} but
      for $q=1$, $\chi_1=\chi_2=+0.9695$ and only the inspiral
      portion. This NR waveform was \emph{not} used to calibrate the
      adjustable parameters $d_{\text{SO}}$ and $d_{\text{SS}}$. Also,
      in the aligned case our prototype EOB model only covers
      $\chi_{1,2} \lesssim 0.7$. Note that in this plot we do not
      include spinning NQC corrections in our EOB waveform.  Alignment
      between the NR and EOB waveforms was performed using
      Eq.~\eqref{alignment}, with $t_{1} = 1170\,M$ and $t_{2} =
      2790\, M$.}
  \end{center}
\end{figure}

For the case $\chi_1=\chi_2=+0.9695$, which is outside the domain of
validity of our prototype EOB model, we cannot successfully run the
NQC iterations, since the NQC corrections are so large that they cause
a divergent sequence of NQC coefficients. Nonetheless, we deem it
interesting to generate the EOB inspiral-plunge waveform where only
the nonspinning NQC coefficients $a_i^{h_{22}}$ ($i=1,2,3$) and
$b_i^{h_{22}}$ ($i=1,2$) are used and compare it to the NR
waveform. In Fig.~\ref{fig:ExtUU} we show how our waveform
performs. We notice that the NR waveform is very long, almost 50 GW
cycles.  The phase difference between the EOB and NR waveforms is
smaller than $0.04$ rads over the first 20 GW cycles, and then grows
up to $0.18$ rads during the subsequent 10 GW cycles and it becomes
$0.9$ rads when 10 GW cycles are left before merger. The fractional
amplitude difference is only $3\%$ when 10 GW cycles are left before
merger.

It is worth emphasizing that although our prototype model is not yet
able to generate merger-ringdown waveforms for spins larger than
$+0.7$, nevertheless, as the comparison with the nearly extremal case
$\chi_1=\chi_2=+0.9695$ has proven, the Hamiltonian of Refs.
~\cite{Barausse:2009xi, Barausse:2011ys} and the resummed flux of
Refs.~\cite{DIN, Pan2010hz} can evolve the EOB dynamics in this highly
relativistic case beyond the orbital-frequency's peak, until $r
\approx 1.9M$, without encountering unphysical features. This suggests
that relevant strong-field effects are well grasped by the EOB
dynamics and waveforms~\cite{Barausse:2009xi, Barausse:2011ys, DIN,
  Pan2010hz}, at least as far as the NR runs used in this paper are
concerned.  Moreover, the large amplitude difference causing the NQC
iteration to break down for large, positive spins was already observed
in Refs.~\cite{Pan2010hz, Barausse:2011kb} where it was pointed out
that it is important to improve the modeling of spin effects in the
EOB waveform amplitude. Finally, as observed above, the breaking down
of the NQC procedure in this highly relativistic case, although not a
problem in principle if higher-order spin-orbit terms were known in
the factorized waveforms, is due to the fact that the peak of the EOB
(2,2) mode occurs too early in the orbital evolution where
non-quasicircular orbit effects are still negligible.


\section{Conclusions}
\label{sec:concl}

Using the EOB spin Hamiltonian in Refs.~\cite{Barausse:2009xi,
  Barausse:2011ys}, the factorized waveforms in Refs.~\cite{DIN,
  Pan2010hz}, and the adjustable parameters in
Table~\ref{tab:adjparams}, we have developed a prototype EOB model for
non-precessing spinning black-hole binaries that can be used for
detection purposes in LIGO and Virgo searches and employed for future
calibrations~\cite{NRARwebsite}. The prototype model is built by first
calibrating the EOB adjustable parameters against five nonspinning
waveforms with mass ratios $q = 1, 1/2, 1/3, 1/4, 1/6$ and two
equal-mass, equal-spin NR waveforms with moderate spins. Then, those
results, at the interface with NR, are combined with recent results at
the interface with black-hole perturbation
theory~\cite{Barausse:2011kb}. The resulting prototype EOB model
interpolates between calibrated points in the binary parameter space,
and generates inspiral-merger-ringdown waveforms with any mass ratio
and individual spin magnitudes $-1\leq \chi_i \lesssim 0.7$.  This EOB
model has been implemented in the freely available LIGO Algorithm
Library (LAL)~\cite{LAL} with the model name
``SEOBNRv1''.\footnote{Two nonspinning EOB models are also available
  in LAL, ``EOBNRv1'' and ``EOBNRv2'', which were calibrated to NR
  waveforms in Refs.~\cite{Buonanno2007, Pan:2011gk}.}

We found that the EOB waveforms generated with the prototype model
agree with the NR waveforms used to calibrate them within $\roughly
0.1$ rads at merger for the nonspinning sector, and within $\roughly
0.15$ rads at merger for the spinning sector. In terms of amplitude
differences at merger, both nonspinning and spinning runs agree to
within $5\%$.  The $\MM$s for Advanced LIGO computed by maximizing only
with respect to the initial phase and time are always smaller than
0.003 for binaries with total masses between $20 M_\odot$ and $200
M_\odot$.

We also compared the prototype EOB model to two equal-mass, equal-spin
NR waveforms of black holes with nearly extremal spins, notably
$\chi_i = -0.94905, +0.9695$.  Those NR waveforms were not part of the
original set of waveforms used to calibrate the EOB model. We found
that for the anti-aligned case the prototype EOB model performs quite
well for detection purposes, with $\MM$s smaller than $0.003$ without
maximizing over the binary parameters, but only on initial phase and
time.  In the aligned case, which is highly relativistic due to a spin
as large as $+0.9695$ (outside the range of validity of our prototype
model), we compared the inspiral-plunge waveform for 40 GW cycles and
found a dephasing of $\roughly 0.8$ rad. During the last 10 GW cycles
before merger the dephasing grows up to several radians.  This
non-satisfactory performance during plunge and merger for large,
positive spins is not surprising. In our prototype spin EOB model the
factorized modes~\cite{Pan2010hz} used in the radiation-reaction force
generate spin couplings in the GW energy flux at a PN order much lower
than what is known today.  In fact, the GW energy flux is currently
known through 3PN order in the spin-orbit
sector\footnote{Reference~\cite{Lovelace:2011nu} found that the tail
  spin-orbit terms in the energy flux at 3PN order dominate all the
  other spin-orbit contributions and improve the agreement with NR
  waveforms.}~\cite{Blanchet:2011zv} and 2PN order in the spin-spin
sector. However, the $-2$ spin-weighted spherical harmonics that are
used to build the factorized waveforms employed in this paper are
known only through 1.5PN order in the spin-orbit
sector~\cite{Arun:2009}.  Moreover, the performance we found for large
spin values and prograde orbits confirms what was already found in
Ref.~\cite{Barausse:2011kb}, where EOB waveforms in the test-particle
limit could be calibrated to Teukolsky-type waveforms only up to a
Kerr spin value of $\roughly +0.7$. For larger spin values, the
factorized waveforms start deviating from the exact ones even before
reaching the ISCO~\cite{Pan2010hz, Barausse:2011kb}.

The prototype spin EOB model can be improved in the future in
different directions. First, the choice of the spin EOB adjustable
parameters done in Sec.~\ref{sec:EOB-model} was rather arbitrary and
assumed that all gauge parameters that enter the spin EOB conservative
dynamics are zero. Of course, it would have been difficult to carry
out a more sophisticated study in this work considering that we had at
our disposal only two equal-mass, equal-spin NR waveforms.  When
several more spin NR waveforms will be available, the spin EOB
parameters (together with the nonspinning ones) should be explored
and calibrated simultaneously against all the available NR
waveforms. Second, it is urgent to compute higher-order PN spin-orbit
terms in the -2 spin weighted spherical harmonics and in the
factorized modes, thus making the EOB spin model reliable also for
large, positive spins, i.e., for $\chi_i > 0.7$. Third, the spin EOB
Hamiltonian at 3.5PN order used in this paper predicts for large,
positive spins that the position of the peak of the EOB
orbital-frequency varies non-monotonically as function of the spin and
lies in a region which is not very relativistic. It would be important
to correct this behavior calibrating the gauge parameters present in
the spin EOB Hamiltonian. Fourth, recent results in
Refs.~\cite{Damour:2009sm, Barack:2010ny, LeTiec:2011ab,
  LeTiec:2011dp} at the interface between PN theory and the self-force
formalism, have allowed Ref.~\cite{Barausse:2011dq} to compute the
nonspinning EOB potentials at all orders in PN theory, but linear in
the symmetric mass ratio $\nu$. These new results will be incorporated
in the future to improve the nonspinning conservative dynamics of the
prototype EOB model, and will be extended to include spin effects.

\begin{acknowledgments}
  E.B., A.B., Y.P., and A.T. acknowledge support from NSF Grants
  No. PHY-0903631.  A.B. also acknowledges support from NASA grant
  NNX09AI81G.  T.C., G.L., M.B., and M.S. are supported in part by
  grants from the Sherman Fairchild Foundation to Caltech and Cornell,
  and from the Brinson Foundation to Caltech; by NSF Grants
  No. PHY-0601459 and No. PHY-0652995 at Caltech; by NASA Grant
  NNX09AF97G at Caltech; by NSF Grants No. PHY-0652952 and
  No. PHY-0652929 at Cornell; and by NASA Grant No. NNX09AF96G at
  Cornell.  H.P. gratefully acknowledges support from the NSERC of
  Canada, from Canada Research Chairs Program, and from the Canadian
  Institute for Advanced Research.
\end{acknowledgments}

\appendix

\section{Explicit expressions of the factorized modes}
\label{sec:AppendixFactModes}

Using results from Refs.~\cite{Damour2009a, Buonanno:2009qa,
  Pan:2009wj, Pan:2011gk}, we write here the explicit expressions of
the factorized modes employed in Sec.~\ref{sec:EOB-wave}. Even though
we calibrated only the (2,2) mode, we will provide expressions for all
the modes up to $\ell=8$, because they enter the computation of the
energy flux in Eq.~\eqref{Edot}.

The terms $h_{\ell m}^{(N,\epsilon)}$ in Eq.~\eqref{hlm} are the
Newtonian modes. They read
\begin{equation}\label{hlmNewt}
  h_{\ell m}^{(N,\epsilon)}=\frac{M\nu}{\cal{R}}\, n_{\ell
    m}^{(\epsilon)}\, c_{\ell+\epsilon}(\nu)\, V^{\ell}_\Phi\,
  Y^{\ell-\epsilon,-m}\,\left(\frac{\pi}{2},\Phi\right)\,,
\end{equation}
where $\cal{R}$ is the distance from the source; the $Y^{\ell
  m}(\Theta,\Phi)$ are the scalar spherical harmonics; we use
$V^{\ell}_\Phi = v_\Phi^{\ell+\epsilon}$ with
\begin{equation}
  v_\Phi= r_{\Omega} \hat{\Omega} = \hat{\Omega}
  \left(\left.\frac{\partial \hat{H}_{\text{real}}}{\partial
        p_\Phi}\right|_{p_r=0}\right)^{-2/3}\,,
\end{equation}
where $p_{\Phi}\equiv|\vvr \times \vp |$. The functions $n_{\ell
  m}^{(\epsilon)}$ and $c_{\ell+\epsilon}(\nu)$ in Eq.~\eqref{hlmNewt}
read
\begin{subequations}
  \label{n0n1}
  \begin{align}
    n^{(0)}_{\ell m} &= (i\, m)^\ell\frac{8\pi}{(2\ell+1)!!}\sqrt{\frac{(\ell+1)(\ell+2)}{\ell(\ell-1)}}\,, \label{nlmeven} \\
    n^{(1)}_{\ell m} &= -(i\, m)^\ell\frac{16\pi
      i}{(2\ell+1)!!}\sqrt{\frac{(2\ell+1)(\ell+2)(\ell^2-m^2)}{(2\ell-1)(\ell+1)\ell(\ell-1)}}\,,
    \label{nlmodd}
  \end{align}
\end{subequations}
and
\begin{multline}
  \label{cl}
  c_{\ell+\epsilon}(\nu) = \left(\frac{1}{2}-\frac{1}{2}\sqrt{1-4\nu}\right)^{\ell+\epsilon-1} \\
  +(-1)^{\ell+\epsilon}\left(\frac{1}{2}+\frac{1}{2}\sqrt{1-4\nu}\right)^{\ell+\epsilon-1}\,.
\end{multline}
The function $\hat{S}_\text{ eff}^{(\epsilon)}$ in Eq.~\eqref{hlm} is
an effective source term that in the circular-motion limit contains a
pole at the EOB light ring. It is given in terms of the EOB dynamics
as
\begin{equation}
  \hat{S}_\text{eff}^{(\epsilon)}(r, p_{r^*}, p_\Phi, \mathbf{S}_1, \mathbf{S}_2) =
  \begin{cases}
    \hat{H}^\text{eff}(r, p_{r^*}, p_\Phi, \mathbf{S}_1, \mathbf{S}_2)\,, & \epsilon = 0\,, \\
    \hat{L}_\text{eff}=p_\Phi\, v_\Omega\,, & \epsilon = 1\,,
  \end{cases}
\end{equation}
where $v_{\Omega}=\hat{\Omega}^{1/3}$. The factor $T_{\ell m}$ in
Eq.~\eqref{hlm} resums the leading order logarithms of tail effects,
it reads
\begin{equation}\label{eq:tailterm}
  \begin{split}
    T_{\ell m} &=\frac{\Gamma(\ell+1-2i\, m\, H_\text{real}\, \Omega)}
    {\Gamma(\ell+1)}\, \exp \left[ \pi\, m\,\Omega\, H_\text{real}
    \right] \\ & \qquad \times \exp \left[ 2i\, m\,\Omega\,
      H_\text{real}\, \log(2\, m\,\Omega\, r_0) \right]~,
  \end{split}
\end{equation}
where $r_0=2M/\sqrt{e}$~\cite{Pan2010hz}.

In what follows we define
\begin{subequations}
  \begin{align}
    \delta m \equiv \frac{m_1 - m_2}{M},\\
    \chi_S \equiv \frac{\chi_1+\chi_2}{2}, \\
    \chi_A \equiv \frac{\chi_1-\chi_2}{2}.
  \end{align}
\end{subequations}
Also we use $\mathrm{eulerlog}_m(v_\Omega^2)\equiv\gamma_E + \log 2 +
\log m + 1/2\log v_\Omega^2$, with $\gamma_E$ being the Euler
constant.
\begin{widetext}
We noticed that for even $m$ the $\rho_{\ell m}$'s with spin contributions of 
Ref.~\cite{Pan2010hz} are ill-defined when $\delta m \rightarrow 0$. Thus, in this paper, 
for $m= 1, 3$ and $\ell \leq 4$, we replace the factor $\left(\rho_{\ell m}\right)^{\ell}$ in Eq.~(\ref{hlm}) with the nonspinning 
(NS) limit of $\left(\rho_{\ell m}\right)^{\ell}$ plus the spinning (S) part of the $f_{\ell m}$'s of Ref.~\cite{Pan2010hz}. More explicitly, 
the modes we used read~\cite{DIN, Pan2010hz}
  \begin{subequations}
    \label{rho2}
    \begin{align}
      \rho_{22}&=1+\left(\frac{55\,\nu}{84}-\frac{43}{42}\right) v_\Omega^2 - \frac{2}{3} \left[ \chi_S(1-\nu)+\chi_A\,{\delta m}\right] v_\Omega^3 + \left(\frac{19\, 583\,\nu^2}{42\, 336}-\frac{33\, 025\,\nu}{21\, 168}-\frac{20\, 555}{10\, 584}\right) v_\Omega^4 \notag\\
      &\quad+ \left(\frac{10\, 620\, 745\,\nu^3}{39\, 118\, 464}-\frac{6\, 292\, 061\,\nu^2}{3\, 259\, 872}+\frac{41\,\pi^2\,\nu}{192}-\frac{48\, 993\, 925\,\nu}{9\, 779\, 616}-\frac{428\,\text{eulerlog}_2(v_\Omega^2)}{105}+\frac{1\, 556\, 919\, 113}{122\, 245\, 200}\right)\, v_\Omega^6\notag\\
      &\quad + \left(\nu \rho_{22}^{(4)}+\frac{9\,
          202\,\text{eulerlog}_2(v_\Omega^2)}{2\, 205}-\frac{387\,
          216\, 563\, 023}{160\, 190\, 110\, 080}\right)\, v_\Omega^8
      \quad+
      \left(\frac{439\, 877\,\text{eulerlog}_2(v_\Omega^2)}{55\, 566}-\frac{16\, 094\, 530\, 514\, 677}{533\, 967\, 033\, 600}\right)\, v_\Omega^{10}\,, \label{rho22}\\
      \rho^{L \,\rm NS}_{21}&=1 + \left(\frac{23\,\nu}{84}-\frac{59}{56}\right)v_\Omega^2 +\left(\frac{617\,\nu^2}{4\, 704}-\frac{10\, 993\,\nu}{14\, 112}-\frac{47\, 009}{56\, 448}\right) v_\Omega^4\notag\\
      &\quad+ \left(\frac{7\, 613\, 184\, 941}{2\, 607\, 897\, 600}-\frac{107\,\text{eulerlog}_1(v_\Omega^2)}{105} \right)\, v_\Omega^6 + \left(\frac{6\, 313\,\text{eulerlog}_1(v_\Omega^2)}{5\, 880}-\frac{1\, 168\, 617\, 463\, 883}{911\, 303\, 737\, 344}\right)\, v_\Omega^8 \notag\\
      &\quad + \left(\frac{5\, 029\,
          963\,\text{eulerlog}_1(v_\Omega^2)}{5\, 927\,
          040}-\frac{63\, 735\, 873\, 771\, 463}{16\, 569\, 158\,
          860\, 800}\right)\, v_\Omega^{10} \label{rho21} \,,
    \end{align}
  \end{subequations}
  where $\rho_{22}^{(4)}$ is a nonspinning EOB adjustable parameter,
  which is determined through the calibration of the nonspinning NR
  waveforms,

  \begin{subequations}
    \label{rho3}
    \begin{align}
      \rho^{\rm NS}_{33}&=1+\left(\frac{2\,\nu}{3}-\frac{7}{6}\right) v_\Omega^2 
+\left(\frac{149\,\nu^2}{330}-\frac{1\, 861\,\nu}{990}-\frac{6\, 719}{3\, 960}\right) v_\Omega^4 \notag \\
      &\quad + \left(\frac{3\, 203\, 101\, 567}{227\, 026\,
          800}-\frac{26\,\text{eulerlog}_3(v_\Omega^2)}{7}\right)\,
      v_\Omega^6 +
      \left(\frac{13\,\text{eulerlog}_3(v_\Omega^2)}{3}-\frac{57\,
          566\, 572\, 157}{8\, 562\, 153\, 600}\right)\, v_\Omega^8
      \label{rho33} \,,\\
      \rho^L_{32}&=1-\frac{4 \nu}{3(3\nu-1)}\chi_S  v_\Omega+\frac{320\,\nu^2-1\, 115\,\nu+328}{270\,(3\,\nu-1)} v_\Omega^2  \notag \\
      &\quad +\frac{3\, 085\, 640\,\nu^4-20\, 338\, 960\,\nu^3-4\, 725\, 605\,\nu^2+8\, 050\, 045\,\nu-1\, 444\, 528}{1\, 603\, 800\,(1-3\,\nu)^2} v_\Omega^4 \notag \\
      &\quad+ \left(\frac{5\, 849\, 948\, 554}{940\, 355\,
          325}-\frac{104\,\text{eulerlog}_2(v_\Omega^2)}{63}\right)\,
      v_\Omega^6 + \left(\frac{17\,
          056\,\text{eulerlog}_2(v_\Omega^2)}{8\, 505}-\frac{10\,
          607\, 269\, 449\, 358}{3\, 072\, 140\, 846\, 775}\right)\,
      v_\Omega^8
      \label{rho32} \,, \\
      \rho^{\rm NS}_{31}&=1-\left(\frac{2\,\nu}{9}+\frac{13}{18}\right) v_\Omega^2
+ \left(-\frac{829\,\nu^2}{1\, 782}-\frac{1\, 685\,\nu}{1\, 782}+\frac{101}{7\, 128}\right) v_\Omega^4 \notag\\
      &\quad+ \left(\frac{11\, 706\, 720\, 301}{6\, 129\, 723\,
          600}-\frac{26\,\text{eulerlog}_1(v_\Omega^2)}{63}\right)\,
      v_\Omega^6 +
      \left(\frac{169\,\text{eulerlog}_1(v_\Omega^2)}{567}+\frac{2\,
          606\, 097\, 992\, 581}{4\, 854\, 741\, 091\, 200}\right)\,
      v_\Omega^8 \label{rho31} \,,
    \end{align}
  \end{subequations}
  
  \begin{subequations}
    \label{rho4}
    \begin{align}
      \rho_{44}&=1+\frac{2\, 625\nu^2-5\, 870\,\nu+1\, 614}{1\, 320\,(3\,\nu-1)} v_\Omega^2 -\frac{1}{15(1-3\nu)} \left[(42 \nu^2-41\nu+10)\chi_S + (10-39\nu) {\delta m}\,\chi_A\right]  v_\Omega^3\notag\\
      &\quad+\frac{1\, 252\, 563\, 795\,\nu^4-6\, 733\, 146\, 000\,\nu^3-313\, 857\, 376\,\nu^2+2\, 338\, 945\, 704\,\nu-511\, 573\, 572}{317\, 116\, 800\,(1-3\,\nu)^2} v_\Omega^4 \notag\\
      &\quad+
      \left(\frac{16\, 600\, 939\, 332\, 793}{1\, 098\, 809\, 712\, 000}-\frac{12\, 568\,\text{eulerlog}_4(v_\Omega^2)}{3\, 465}\right)\, v_\Omega^6\,,\label{rho44}\\
      \rho^{L \, \rm NS}_{43}&=1+\frac{160\,\nu^2-547\,\nu+222}{176\,(2\,\nu-1)}
      v_\Omega^2-\frac{6\, 894\, 273}{7\, 047\, 040} v_\Omega^4 + \left(\frac{1\, 664\, 224\, 207\, 351}{195\, 343\,
          948\, 800}-\frac{1\,
          571\,\text{eulerlog}_3(v_\Omega^2)}{770}\right)\, v_\Omega^6
      \,, \label{rho43}\\
      \rho_{42}&=1+\frac{285\,\nu^2-3\, 530\,\nu+1\, 146}{1\, 320\,(3\,\nu-1)} v_\Omega^2 - \frac{1}{15(1-3\nu)} \left[(78\nu^2 - 59\nu + 10)\chi_S +(10-21\nu){\delta m}\,\chi_A \right] v_\Omega^3 \notag \\
      &\quad+\frac{-379\, 526\, 805\,\nu^4-3\, 047\, 981\, 160\,\nu^3+1\, 204\, 388\, 696\,\nu^2+295\, 834\, 536\,\nu-114\, 859\, 044}{317\, 116\, 800\,(1-3\,\nu)^2} v_\Omega^4 \notag\\
      &\quad+ \left(\frac{848\, 238\, 724\, 511}{219\, 761\, 942\,
          400}-\frac{3\, 142\,\text{eulerlog}_2(v_\Omega^2)}{3\,
          465}\right)\, v_\Omega^6
      \,,\label{rho42}\\
      \rho^{L \, \rm NS}_{41}&=1 +\frac{288\,\nu^2-1\,
        385\,\nu+602}{528\,(2\,\nu-1)}
      v_\Omega^2-\frac{7\, 775\, 491}{21\, 141\, 120} v_\Omega^4 + \left(\frac{1\, 227\, 423\, 222\, 031}{1\, 758\, 095\,
          539\, 200}-\frac{1571\,\text{eulerlog}_1(v_\Omega^2)}{6\,
          930}\right)\, v_\Omega^6 \,,\label{rho41}
    \end{align}
  \end{subequations}

  \begin{subequations}
    \label{rho5}
    \begin{align}
      \begin{split}
        \label{rho55}
        \rho_{55}&=1-\frac{512\,\nu^2-1\,
          298\,\nu+487}{390\,(2\,\nu-1)} v_\Omega^2 -\frac{3\, 353\,
          747}{2\, 129\, 400} v_\Omega^4\,,
      \end{split}
      \\
      \begin{split}
        \label{rho54}
        \rho^L_{54}&=1+\frac{33\, 320\,\nu^3-127\, 610\,\nu^2+96\,
          019\,\nu-17\, 448}{13\, 650\,(5\,\nu^2-5\,\nu+1)} v_\Omega^2
        -\frac{16\, 213\, 384}{15\, 526\, 875} v_\Omega^4 \,,
      \end{split}
      \\
      \begin{split}
        \label{rho53}
        \rho_{53}&=1+\frac{176\,\nu^2-850\,\nu+375}{390\,(2\,\nu-1)}
        v_\Omega^2 -\frac{410\, 833}{709\, 800} v_\Omega^4\,,
      \end{split}
      \\
      \begin{split}
        \label{rho52}
        \rho^L_{52}&=1+\frac{21\, 980\,\nu^3-104\, 930\,\nu^2+84\,
          679\,\nu-15\, 828}{13\, 650\,(5\,\nu^2-5\,\nu+1)} v_\Omega^2
        -\frac{7\, 187\, 914}{15\, 526\, 875} v_\Omega^4 \,,
      \end{split}
      \\
      \begin{split}
        \label{rho51}
        \rho_{51}&=1+\frac{8\,\nu^2-626\,\nu+319}{390\,(2\,\nu-1)}
        v_\Omega^2 -\frac{31\, 877}{304\, 200} v_\Omega^4\,,
      \end{split}
    \end{align}
  \end{subequations}
  
  \begin{subequations}
    \begin{align}
      \rho_{66} &= 1+\frac{273\,\nu^3-861\,\nu^2+602\,\nu-106}
      {84\,(5\,\nu^2-5\,\nu+1)} v_\Omega^2 - \frac{1\, 025\, 435}
      {659\, 736} v_\Omega^4 \,, \label{rho66} \\
      \rho^L_{65}& = 1 + \frac{220\,\nu^3-910\,\nu^2+838\,\nu-185}
      {144\,(3\,\nu^2-4\,\nu+1)} v_\Omega^2\,, \label{rho65} \\
      \rho_{64} &= 1 + \frac{133\,\nu^3-581\,\nu^2+462\,\nu-86}
      {84\,(5\,\nu^2-5\,\nu+1)} v_\Omega^2 - \frac{476\, 887}
      {659\, 736} v_\Omega^4 \,, \label{rho64} \\
      \rho^L_{63} &= 1 + \frac{156\,\nu^3-750\,\nu^2+742\,\nu-169}
      {144\,(3\,\nu^2-4\,\nu+1)} v_\Omega^2\,, \label{rho63} \\
      \rho_{62} &= 1 + \frac{49\,\nu^3-413\,\nu^2+378\,\nu-74}
      {84\,(5\,\nu^2-5\,\nu+1)} v_\Omega^2 - \frac{817\, 991}
      {3\, 298\, 680} v_\Omega^4 \,, \label{rho62} \\
      \rho^L_{61} &= 1 + \frac{124\,\nu^3-670\,\nu^2+694\,\nu-161}
      {144\,(3\,\nu^2-4\,\nu+1)} v_\Omega^2\,, \label{rho61}
    \end{align}
  \end{subequations}

  \begin{subequations}
    \begin{align}
      \rho_{77} &= 1 + \frac{1380 \nu ^3-4963 \nu ^2+4246 \nu -906}
      {714 \left(3 \nu ^2-4 \nu +1\right)} v_{\Omega }^2 \,, \\
      \rho^L_{76} &= 1 + \frac{6104 \nu ^4-29351 \nu ^3+37828 \nu
        ^2-16185 \nu +2144} {1666 \left(7 \nu ^3-14 \nu ^2+7 \nu
          -1\right)} v_{\Omega }^2 \,, \\
      \rho_{75} &= 1 + \frac{804 \nu ^3-3523 \nu ^2+3382 \nu -762}
      {714 \left(3 \nu ^2-4 \nu +1\right)} v_{\Omega }^2 \,, \\
      \rho^L_{74} &= 1 + \frac{41076 \nu ^4-217959 \nu ^3+298872 \nu
        ^2-131805 \nu +17756} {14994 \left(7 \nu ^3-14 \nu ^2+7 \nu
          -1\right)} v_{\Omega }^2 \,, \\
      \rho_{73} &= 1 + \frac{420 \nu ^3-2563 \nu ^2+2806 \nu -666}
      {714 \left(3 \nu ^2-4 \nu +1\right)} v_{\Omega }^2 \,, \\
      \rho^L_{72} &= 1 + \frac{32760 \nu ^4-190239 \nu ^3+273924 \nu
        ^2-123489 \nu +16832} {14994 \left(7 \nu ^3-14 \nu ^2+7 \nu
          -1\right)} v_{\Omega }^2 \,, \\
      \rho_{71} &= 1 + \frac{228 \nu ^3-2083 \nu ^2+2518 \nu -618}
      {714 \left(3 \nu ^2-4 \nu +1\right)} v_{\Omega }^2 \,,
    \end{align}
  \end{subequations}
  
  \begin{subequations}
    \begin{align}
      \rho_{88} &= 1+ \frac{3482 - 26778\nu + 64659\nu^2 - 53445\nu^3 + 12243\nu^4}{2736 (-1 + 7 \nu- 14 \nu^2 + 7 \nu^3)} v_\Omega^2\,, \\
      \rho^L_{87} &= 1+ \frac{23478 - 154099\nu + 309498\nu^2 - 207550\nu^3 + 38920\nu^4}{18240 (-1 + 6\nu - 10\nu^2 + 4\nu^3)} v_\Omega^2\,, \\
      \rho_{86} &= 1+ \frac{1002 - 7498\nu + 17269\nu^2 - 13055\nu^3 + 2653\nu^4}{912(-1 + 7\nu - 14\nu^2 + 7\nu^3)} v_\Omega^2\,, \\
      \rho^L_{85} &= 1+ \frac{4350 - 28055\nu + 54642\nu^2 - 34598\nu^3 + 6056\nu^4}{3648 (-1 + 6\nu - 10\nu^2 + 4\nu^3)} v_\Omega^2\,, \\
      \rho_{84} &= 1+ \frac{2666 - 19434\nu + 42627\nu^2 - 28965\nu^3 + 4899\nu^4}{2736(-1 + 7\nu - 14\nu^2 + 7\nu^3)} v_\Omega^2\,, \\
      \rho^L_{83} &= 1+ \frac{20598 - 131059\nu + 249018\nu^2 - 149950\nu^3 + 24520\nu^4}{18240(-1 + 6\nu - 10\nu^2 + 4\nu^3)} v_\Omega^2\,, \\
      \rho_{82} &= 1+ \frac{2462 - 17598\nu + 37119\nu^2 - 22845\nu^3 + 3063\nu^4}{2736(-1 + 7\nu - 14\nu^2 + 7\nu^3)} v_\Omega^2\,, \\
      \rho^L_{81} &= 1+ \frac{20022 - 126451\nu + 236922\nu^2 -
        138430\nu^3 + 21640\nu^4}{18240(-1 + 6\nu - 10\nu^2 + 4\nu^3)}
      v_\Omega^2\,\ab{\,,}
    \end{align}
  \end{subequations}
and 
\begin{subequations}
\begin{align}
f^{L\,\textrm{S}}_{21} &= - \frac{3}{2} \left(\chi_S + \frac{\chi_A}{\delta m}\right) v_{\Omega}\, , \\
f^{\textrm{S}}_{33} &= - \left[\chi_S \left(2 - \frac{5}{2} \nu \right) +\frac{\chi_A}{\delta m} \left(2 - \frac{19}{2} \nu \right)\right] v_\Omega^3\, , \\
f^{L\,\textrm{S}}_{31} &= -\left[ \chi_S \left(2 - \frac{11}{2} \nu \right) + \frac{\chi_A}{\delta m} \left(2 - \frac{13}{2} \nu \right)\right] v_\Omega^3\, , \\
f^{L\,\textrm{S}}_{43} &=  f^{L\,\textrm{S}}_{41} = -\frac{5\nu}{2(2\nu-1)}\left(\chi_S-\frac{\chi_A}{\delta m} \right) v_\Omega\, .
\end{align}
\end{subequations}
Finally, we give the explicit expression of the phase term
  \begin{equation}
    \delta_{22} = \frac{7}{3}
    \left(\hat{\Omega}\,H_{\text{real}}\right) +\frac{428 \pi}{105}
    \left(\hat{\Omega}\,H_{\text{real}}\right)^2 +
    \left(\frac{1712\pi^2}{315}-\frac{2203}{81}\right)
    \left(\hat{\Omega}\,H_{\text{real}}\right)^3 - 24 \nu\,v_\Omega^5.
  \end{equation}
\end{widetext}
\bibliography{References/References}
\end{document}